\documentclass[12pt,preprint]{aastex}
\usepackage{amssymb,amsmath}
\usepackage{hyperref,color}
\newcounter{address}
\newcounter{tableone}
\newcounter{tabletwo}
\definecolor{linkcolor}{rgb}{0,0,0.25}
\hypersetup{
  colorlinks=true,        
  linkcolor=linkcolor,    
  citecolor=linkcolor,    
  filecolor=linkcolor,    
  urlcolor=linkcolor      
}
\newcommand{\eqnnumber}{equation}
\newcommand{\eg}{e.g.}
\newcommand{\ie}{i.e.}

\newcommand{\Hipparcos}{\textit{Hipparcos}}
\newcommand{\Gaia}{\textit{Gaia}}

\newcommand{\gcs}{\textit{Geneva-Copenhagen Survey}}
\newcommand{\gcsabb}{GCS}
\newcommand{\normal}{{\cal N}}

\renewcommand{\vec}[1]{\mathbf{#1}} 

\newcommand{\bb}{\vec{b}}
\newcommand{\cc}{\vec{c}}

\newcommand{\mm}{\vec{m}}
\newcommand{\vv}{\vec{v}}

\newcommand{\ww}{\vec{w}}
\newcommand{\bij}{\bb_{ij}}
\newcommand{\bbij}{\bij}
\newcommand{\cci}{\cc_i}
\newcommand{\eex}{\vec{\hat{x}}}
\newcommand{\eey}{\vec{\hat{y}}}
\newcommand{\eez}{\vec{\hat{z}}}
\newcommand{\eer}{\vec{\hat{r}}}
\newcommand{\mmj}{\mm_j}
\newcommand{\mmk}{\mm_k}
\newcommand{\mmrj}{m_{r,j}}
\newcommand{\vvi}{\vv_i}

\newcommand{\wwi}{\ww_i}

\newcommand{\ten}[1]{\mathbf{#1}} 
\newcommand{\BB}{\ten{B}}
\newcommand{\CC}{\ten{C}}
\newcommand{\UU}{\ten{U}}
\newcommand{\QQ}{\ten{Q}}
\newcommand{\RR}{\ten{R}}
\renewcommand{\SS}{\ten{S}}
\newcommand{\TT}{\ten{T}}
\newcommand{\AAA}{\ten{A}}
\newcommand{\AAAi}{\AAA_i}
\newcommand{\VV}{\ten{V}}

\newcommand{\II}{\ten{I}}
\newcommand{\BBij}{\BB_{ij}}
\newcommand{\CCi}{\CC_i}
\newcommand{\UUi}{\UU_i}
\newcommand{\QQi}{\QQ_i}
\newcommand{\radialproj}{\RR_r}
\newcommand{\tangproj}{\RR_t}
\newcommand{\RRi}{\RR_i}

\newcommand{\SSi}{\SS_i}
\newcommand{\SSt}{\SS_t}
\newcommand{\VVj}{\VV_{\!j}} 
\newcommand{\TTij}{\TT_{ij}}
\newcommand{\TTrj}{T_{r,j}}
\newcommand{\TTik}{\TT_{ik}}
\newcommand{\T}{^{\scriptscriptstyle \top}}   
\newcommand{\alphaj}{\alpha_j}

\newcommand{\qij}{q_{ij}}

\newcommand{\qqj}{q_j}

\newcommand{\alphagauss}{\ensuremath{\alpha}}
\newcommand{\ngauss}{\ensuremath{j}}
\newcommand{\dd}{\mathrm{d}}

\newcommand{\ra}{\ensuremath{\alpha}}
\newcommand{\dec}{\ensuremath{\delta}}
\newcommand{\pmra}{\ensuremath{\mu_{\ra}}}
\newcommand{\pmrastar}{\ensuremath{\mu_{\ra *}}}
\newcommand{\pmdec}{\ensuremath{\mu_{\dec}}}

\newcommand{\parallax}{\ensuremath{\varpi}}

\newcommand{\vrr}{\ensuremath{v_r}}

\newcommand{\ngp}{\textnormal{NGP}}

\newcommand{\rangp}{\ensuremath{\ra_\ngp}}
\newcommand{\decngp}{\ensuremath{\dec_\ngp}}

\newcommand{\degree}{^{\circ}}
\newcommand{\matrixleft}{\left[}
\newcommand{\matrixright}{\right]}
\newcommand{\arcsecs}{\textnormal{as}}
\newcommand{\order}[1]{\ensuremath{\mathcal{O}(#1)}}
\newcommand{\sectionname}{Section}
\newcommand{\vx}{\ensuremath{v_x}}
\newcommand{\vy}{\ensuremath{v_y}}
\newcommand{\vz}{\ensuremath{v_z}}
\newcommand{\EM}{EM}
\newcommand{\eqnname}{equation}
\newcommand{\figuresname}{Figures}
\newcommand{\unitmatrix}{\ensuremath{\ten{I}}}

\newcommand{\AIC}{AIC}
\newcommand{\MDL}{MDL}
\newcommand{\MML}{MML}
\newcommand{\nparam}{\ensuremath{N_{\mbox{{\footnotesize param}}}}}
\newcommand{\totalnparam}{\ensuremath{N_{\mbox{{\footnotesize param,tot}}}}}

\begin{document}

\title{The velocity distribution of nearby stars from \Hipparcos\ data\\
I. The significance of the moving groups}
\author{
Jo~Bovy\altaffilmark{\ref{NYU},\ref{email}}, 
David~W.~Hogg\altaffilmark{\ref{NYU},\ref{MPIA}}, 
and Sam~T.~Roweis\altaffilmark{\ref{Toronto},\ref{Google}}} 
\altaffiltext{\theaddress}{\label{NYU}\stepcounter{address}
Center for Cosmology and Particle Physics, Department of Physics, New York 
University, 4 Washington Place, New York, NY 10003}
\altaffiltext{\theaddress}{\stepcounter{address}\label{email}
To whom correspondence should be addressed: \texttt{jo.bovy@nyu.edu}}
\altaffiltext{\theaddress}{\label{MPIA}\refstepcounter{address}
  Max-Planck-Institut f\"ur Astronomie,
  K\"onigstuhl 17, D-69117 Heidelberg, Germany}
\altaffiltext{\theaddress}{\label{Toronto}\stepcounter{address}
Department of Computer Science, University of Toronto, 
6 King's College Road, Toronto, Ontario, M5S 3G4 Canada}
\altaffiltext{\theaddress}{\label{Google}\stepcounter{address}
Google Inc., Mountain View, CA}

\begin{abstract}
We present a three-dimensional reconstruction of the velocity
distribution of nearby stars ($\lesssim 100$ pc) using a maximum
likelihood density estimation technique applied to the two-dimensional
tangential velocities of stars. The underlying distribution is modeled
as a mixture of Gaussian components. The algorithm reconstructs the
error-deconvolved distribution function, even when the individual
stars have unique error and missing-data properties. We apply this
technique to the tangential velocity measurements from a kinematically
unbiased sample of 11,865 main sequence stars observed by the
\Hipparcos\ satellite.  We explore various methods for validating the
complexity of the resulting velocity distribution function, including
criteria based on Bayesian model selection and how accurately our
reconstruction predicts the radial velocities of a sample of stars
from the Geneva-Copenhagen survey (\gcsabb). Using this very
conservative external validation test based on the \gcsabb, we find
that there is little evidence for structure in the distribution
function beyond the moving groups established prior to the \Hipparcos\
mission. This is in sharp contrast with internal tests performed here
and in previous analyses, which point consistently to maximal
structure in the velocity distribution. We quantify the information
content of the radial velocity measurements and find that the mean
amount of new information gained from a radial velocity measurement of
a single star is significant. This argues for complementary radial
velocity surveys to upcoming astrometric surveys.
\end{abstract}

\section{Introduction}

One of the key goals of Galactic astronomy---or near-field
cosmology---is to understand the structure and evolutionary history of
the Galaxy. Past and ongoing surveys have consistently found that the
structure of the Galaxy is more complex than previously thought, and
it is very likely, both from a theoretical perspective
\citep{ghigna98a,johnston98a,2003MNRAS.339..834H} as well as from an
observational perspective
\citep[\eg,][]{koposov,Koposov:2009ru,2008ApJ...688..277T,2007ApJ...670..313S},
that upcoming surveys will reveal much more complicated structures,
and in much larger quantities, than those that are observed today. In
the euphoria of a discovery, the statistical significance of the
observed complexity is often only briefly touched upon. In order to
make progress, however, it is important to ask the question whether
the model, which can be arbitrarily complex, is warranted by the
observations, or whether the structure in the data simply represents
statistical fluctuations. This last point is not merely pedantic, as
the significance and the reason for complex substructure has important
ramifications for those who create models for the evolution and
dynamical structure of the Galaxy and for those who interpret the
observed substructure in a cosmological context.  In the present set
of papers we set out (i) to address this question of the complexity of
the underlying distribution of an observed sample of stars using
justifiable statistical methods in one specific example---the
distribution of nearby stars in velocity space---and (ii) to assess the
ramifications of the distribution that we recover and its complexity
for questions concerning the structure and dynamics of the Galactic
disk.

The discussion of the complexity of the structure of the Galaxy has
traditionally focused on the number of components needed to give a
good description of the observed distribution of stars in position,
kinematics, metallicity, and age. And while the discussion has shifted
mostly from a question at the beginning of the twentieth century about
the appropriate number of ``drifts'' necessary to describe the
kinematics of stars near the Sun \citep[see
below;][]{kapteyn05a,schwarzschild07a} through a discussion about the
number of distinct stellar populations
\citep{1944ApJ...100..137B,baade1958a,1954AJ.....59..307R,oort58a,schwarzschild58a}
to an argument about the number of components that make up the
large-scale structure of the Galaxy \citep[in particular, the
existence and properties of the ``thick
disk'';][]{1983MNRAS.202.1025G,1980ApJS...44...73B,1984ApJS...55...67B,2008ApJ...684..287I},
this does not mean that the old controversies were ever fully
resolved. Indeed, the history of the debate on the kinematics of stars
in the local neighborhood provides a fine example of the analysis of
the amount of structure, or lack thereof, in the phase-space
distribution of stars and the consequences of the perceived structure
for the fundamental parameters of the Galaxy.

\subsection{The ``transient'' nature of moving groups}

The first account of a common motion of stars appeared over 160 years
ago in the Pleiades \citep{1846AN.....24..213M}. This common motion
and the fact that the motions of stars further away from the center of
the cluster seemed to be faster was then interpreted as indicating
that the center of the Galaxy lies in the Pleiades, more specifically,
that it coincides with Alcyone, the brightest star in the Pleiades
cluster \citep{madler47}. However, it was soon found that this was but
a single instance of a more common phenomenon, as similar centers of
common motion were also found to exist in other areas of the sky, \ie,
in the Hyades cluster and for five of the bright stars in Ursa Major
\citep{proctor69a}.

The existence of the Hyades and the Ursa Major ``moving groups'', as
the groups of stars with a common motion are commonly called, was
later confirmed and established on a firmer footing by the development
of the convergent-point technique for the determination of the
position of a cluster of stars
\citep{boss08a,1909ApJ....30..135H}. These groups, as well as the
Pleiades moving group discovered earlier, have survived further
scrutiny
\citep{1938AJ.....47...49R,1949ApJ...110..205R,1998AJ....115.2384D}
and we also unambiguously detect them in our data sample. After these
successfull applications of this new technique it was widely used to
find other moving groups and many were found in different areas of the
sky---the Perseus moving cluster \citep{1910MNRAS..71...43E}, the 61
Cygni cluster \citep{1911AJ.....27...33B,1912AJ.....27...96R}, the
Scorpio-Centaurus cluster \citep{1913MNRAS..73..492P}, the Vela moving
cluster \citep{1914ApJ....40...43K}, and the Corona Borealis moving
cluster \citep{rasmuson21}. Most of these are now believed to be
spurious \citep[\eg][]{rasmuson21,1940MNRAS.100..574C}. The reasons
for this failure of the convergent-point method to be a reliable
indicator for the existence of moving groups were threefold: (1) it
does not take into account the observational errors of the proper
motions used and assumes that the space velocities of all of the
individual stars in the cluster are exactly the same, which cannot
hold in the presence of observational errors and which is not even
close to true in the ideal case; (2) it assumes that all three of the
components of the space velocity of the stars in the cluster are the
same, but it is known that mixing in the direction out of the disk is
much more efficient than mixing in the plane of the disk, such that
groups of stars can share a similar motion in the components of the
velocity in the plane of the disk much longer than they can do this in
the direction out of the plane (and more precise determinations of the
velocity distribution have shown that the distribution in directions
out of the disk is essentially featureless, see
\figurename~\ref{fig:annotated_veldist}); (3) the tolerances with
which a star has to meet the convergent-point test for cluster
membership were gradually loosened, in order to deal with the
complications described in the previous two points, leading to the
inclusion of more and more spurious members. Thus, wrong conclusions
were drawn about the amount of structure in the velocity distribution
after the application of a technique that could not reliably determine
the structure of the underlying distribution from an observed sample
of stars, and that was stretched beyond where it was applicable in the
best cases.

During the latter half of the twentieth century the business of
claiming the existence of new moving groups---followed by calling
their reality into question---remained in full swing. Eggen in
particular was prolific in finding new moving groups and much of the
discussion about their existence and members focused on their having a
single stellar population, since all of the moving groups at the time
were thought to be the remnants of disintegrating star clusters. About
a dozen new moving groups were found by Eggen
\citep{1958MNRAS.118..154E,1959MNRAS.119..255E,1959Obs....79..182E,1959Obs....79...88E,1964RGOB...84..111E,1965Obs....85..191E,1969PASP...81..553E,1971PASP...83..271E,1971PASP...83..251E,1978ApJ...222..203E}
and different results for their chemical homogeneity and ages were
found by different groups
\citep[\eg,][]{1970PASP...82...99E,1971MNRAS.153..171W,1975PASP...87...17B,1979LIACo..22..355T,1983AJ.....88..813E,1983MNRAS.204..841M,1986AJ.....92..910E,1988Ap&SS.145...61P}. Most
of these groups are not recovered in more recent analyses. The
existence of the Hyades and Ursa Major groups was established on a
firmer footing by ever growing samples of member stars
\citep{1968SvA....12..279O,1970SvA....13..934O,1970PASP...82...99E,1975PASP...87...17B,1993AJ....105..226S},
although their interpretation as dispersing star clusters was
repeatedly called into question, especially in the case of the Hyades
moving group
\citep{1966Sci...151.1487W,1968PASP...80..578B,1971MNRAS.153..171W,1975PASP...87...17B,1987AJ.....93..920S,1988ApJ...332..410B}. Both
of these groups were believed to contain significant fractions of the
stars in the Solar neighborhood, but it was only after the advent of
complete samples of stars with accurate astrometry with \Hipparcos\
that this question could be studied in detail \citep[with the
exception of][]{1985A&A...145..331G,1990A&A...236...95G}.

\subsection{The \Hipparcos\ era}

The astrometric ESA space mission \Hipparcos, which collected data
over a 3.2 year period around 1990, provided for the first time an
all-sky catalogue of absolute parallaxes and proper motions with
micro-arcsecond precision \citep{ESA97a}. Different complete samples
of stars were extracted from the $\sim\!100,000$ star catalogue, and
many different methods were used to determine the distribution of
these stars in velocity space and its overdensities
\citep{1997A&A...318...29C,1997ESASP.402..519F,1998AJ....115.2384D,1998A&A...340..384C,1999A&AS..135....5C,1999A&A...341..427A,1999A&A...341...86B,1999MNRAS.308..731S}. The
picture that emerged from these various reconstructions of the
velocity distribution was startling: the distribution was revealed to
be extremely structured, with many, if not most, of the stars part of
large associations of stars, in particular, the Hyades, Pleiades and
Ursa Major moving groups \citep{1998AJ....115.2384D}. In addition to
the confirmation of the moving groups, more complex structures such as
several almost parallel branches and sharp edges in the distribution
in the Galactic plane were also observed
\citep{1999MNRAS.308..731S}. Over the years, many radial velocities
measurements have been made, most notably as part of the \gcs\
\citep{2004A&A...418..989N}, which has led to a confirmation of many
of these structures \citep{2005A&A...430..165F}, and a proliferation
of new moving groups has been found in the combination of the data
sets \citep{2008A&A...490..135A,2009ApJ...692L.113Z}. However, the
existence of these new moving groups has not been convincingly shown
to date.

The most popular method by far in recent years to determine the
 velocity distribution is the kernel density estimation technique
 \citep{Silverman86a} and the related wavelet analysis technique for
 identifying moving groups \citep[\eg,][]{1990A&A...227..301S}. This
 technique has been applied to the \Hipparcos\ data at various degrees
 of sophistication. In the simplest form this method basically amounts
 to a smoothed histogram of the data in which each data point is
 replaced by a Gaussian probability distribution (the so-called
 ``kernel'') with some fixed width around the observed value
 \citep{1997A&A...318...29C}. When a kernel with a vanishing volume is
 chosen this technique is known as wavelet analysis, in which the
 kernel is a wavelet that picks out overdensities in the data
 \citep[\eg, a ``Mexican hat'' wavelet,][]{1989wtfm.conf..239M}. When
 applied with a fixed width $\sigma$ this method naturally picks out
 the slightest overdensities of scale $\sigma$ in the observed
 distribution. Therefore, it is unsurprising that setting $\sigma
 \approx 1$ km s$^{-1}$ turns up a large number of overdensities of
 exactly this size \citep{2009arXiv0901.3503F}, which are then
 tentatively called new moving groups \citep{2009ApJ...692L.113Z}, and
 finds substructure in the classical moving groups. Given the
 measurement uncertainties and the spatial extent of the \Hipparcos\
 sample---even a ``cold'' moving group on a circular orbit spread out
 over $\Delta x\approx 100$ pc will create elongated structures in the
 velocity distribution with a size $\approx \Delta x/R_\odot \times
 220$ km s$^{-1}$ $\approx$ 2.5 km s$^{-1}$---structure on these
 scales is unlikely to be real, unless they can be localized in
 position space as well. Slightly more sophisticated techniques set
 the smoothing scale in an optimal way given the kernel, the
 dimensionality of the distribution, and the number of data points
 \citep{1990A&A...235...94C,1999A&A...341..427A}. This theoretically
 optimal value, however, generally leads to scales $\lesssim 1$ km
 s$^{-1}$, such that the smoothing scale problem persists.

More sophisticated analyses realize that the measurement uncertainties
limit the scales on which structure can be detected and that setting
the scale parameter means setting the scale of the detected
structure. Using more realistic values for the scale ($\sigma \approx
5$ km s$^{-1}$) and considering multiple scales returns merely the
classical moving groups from a subset of the \Hipparcos\ data
\citep{1997ESASP.402..519F}. Adaptive kernel methods are techniques in
which the smoothing scale is allowed to vary from data point to data
point, and a well-defined procedure for iteratively setting these
different scales based on the reconstructed distribution and a
leave-one-out cross validation procedure exists
\citep{Silverman86a}. Using a sample of 4,000 \Hipparcos\ stars the
optimal overall smoothing scale was found to be $\approx 11$ km
s$^{-1}$ \citep{1999MNRAS.308..731S} and the reconstructed
distribution showed only a few peaks.

Sophisticated multi-scale methods borrowed from astronomical image
analysis have also been used a few times to determine the velocity
field. These multiscale methods, such as the \emph{{\`a} trous} method
\citep{Starck06}, smooth the observed density on different scales
allowing the study of the velocity distribution on various scales
\citep{1998A&A...340..384C,1999A&AS..135....5C}. These multiscale
methods can be combined with denoising techniques, which filter the
wavelet coefficients according to their significance assuming a prior
distribution on the coefficients \citep[\eg, Wiener
filtering,][]{Starck06}. Since the scales are still set by hand, if
set to a small scale these methods still naturally find structures on
the smallest scale analysed, leading to a large abundance of structure
in the observed velocity distribution \citep{2008A&A...490..135A}.

The advantages of these kernel density and wavelet analysis techniques
are that they are conceptually simple, non-parametric, and
computationally inexpensive, as they behave for the most part as
\order{N}, where $N$ is the number of data points, multiplied by the
number of scales for multiscale methods. One disadvantage of these
techniques is that they only work in the case of complete data, \ie,
data with all of the dimensions measured. Since \Hipparcos\ did not
measure the radial velocities of the survey stars, these techniques
could not be applied to the \Hipparcos\ catalogue by itself, instead
they had to take small subsamples of the catalogue for which full
phase space information was available or wait until the arrival of
radial velocities for a selected number of stars. The main
disadvantage, however, is that in the presence of sizeable measurement
errors, as is the case in the determination of the velocity
distribution from \Hipparcos\ data for which the $\sim\!10$\,percent
parallax uncertainties give rise to $\gtrsim\!10$\,percent velocity
errors, the kernel density estimate reconstructs the \emph{observed}
distribution and not the \emph{underlying} distribution, \ie, they do
not reconstruct the distribution you would find if you had ``good''
data, that is, data with vanishingly small uncertainties and all
dimensions measured. The kernel density estimate and wavelet analysis
techniques do not take into account the individual uncertainty
properties of the data points, at best they let the overall
uncertainty scale guide the choice of smoothing scale. In order to
reconstruct the underlying velocity distribution it is necessary to
convolve a model for the underlying distribution with the
uncertainties of the individual stars and compare the resulting
distribution with the observed distribution. When attempting to
reconstruct the velocity distribution from incomplete data this is the
approach that must be adopted.

\citet{1998AJ....115.2384D} determined the velocity distribution from
\Hipparcos\ data alone. The underlying distribution was pixelized and
the value in each pixel was obtained by a maximum penalized likelihood
method. In order to compare the model with the data Dehnen used an
uncertainty model that consisted of an infinite variance in the radial
direction (the unobserved direction) and a zero variance in the
tangential direction, as he believed that taking into account the
individual errors in the tangential velocities was unnecessary given
the sample size. This simplification leads to an unbounded likelihood
function, which is why the function to be maximized had to include a
penalty functional which penalizes rough distribution functions. Given
the scarcity of the data and the large number of pixels necessary in
three dimensions, Dehnen found that the \Hipparcos\ sample only
samples $\sim\!20$ pixels in each direction for a three-dimensional
reconstruction, and therefore he could only reliably reconstruct
two-dimensional projections of the velocity field. This means that
associating structures in different projections poses somewhat of a
problem. The validation of the resulting velocity distribution
consisted of comparing the reconstructed distribution for different
subsamples of the data and the only reliably identifiable structures
corresponded to the classical moving groups and a thick disk moving
group, the Arcturus stream. Although Dehnen claimed that the various
wiggles in the contours of the reconstructed velocity distribution
correspond to real features, and some of these features are indeed
present in later analyses, one has to remember that all of the samples
derived from the \Hipparcos\ catalogue overlap significantly and that
therefore cross-validation between different analyses does not
necessarily mean cross-validation between different samples. Dehnen's
was the first approach to utilize all of the relevant \Hipparcos\ data
in determining the velocity distribution and he showed that, through
the use of a well-defined, justifiable algorithm designed to cope with
missing data, the large number of \Hipparcos\ measurements could give
the best determination of the velocity field yet.

The first, and until now only, attempt to model the distribution of
stars in a phenomenological way again confirmed the classical moving
groups and found no evidence for new structure
\citep{2005A&A...430..165F}. A mixture of different base groups was
fitted by using a maximum likelihood method which correctly accounted
for the observational uncertainties and the selection function
\citep{1996A&AS..117..405L}. Each base groups consisted of a spatial
distribution that was an exponential disk with a scale-height and was
uniform in the Galactic plane, a velocity distribution that was
Gaussian, a Gaussian luminosity distribution, and a correction for
interstellar absorption. Six base groups were found to be necessary to
provide an acceptable fit to the data by using a likelihood test, the
Wilk's test \citep{1990ebua.conf..407S}. This method, while being
objective and well-justified, has some drawbacks. It assumes that
groups in velocity space also share characteristics in other
observables and it is highly parametric, since each clump in velocity
space has very definite spatial and luminosity distributions
associated with it. The model for the velocity distribution itself is
also highly restrictive, as not only does it assume that each clump in
velocity space has a Gaussian distribution, this Gaussian distribution
is further constrained to be aligned with the Galactic plane (a vertex
deviation is kept as a free parameter).

\subsection{Now, why are we so special?}

In this paper we use a technique which combines all of the good points
of the techniques described above to reconstruct the velocity
distribution. In \sectionname\ \ref{sec:model} we describe our model
of the underlying distribution function, which consists of a mixture
of three-dimensional Gaussian distributions, which are left completely
free. We keep the number of Gaussian components as a free parameter
such that this technique is non-parametric in the sense that a mixture
of a sufficiently large number of normal distributions can fit any
distribution function desired and we let the data decide the
complexity of the model instead of setting the number of components by
hand. Since our confrontation of the model with the data includes
convolving the model with the individual uncertainties of the
observations, this technique correctly takes care of measurement
uncertainties and consequently can deal with missing data. This allows
us to use one large sample of stars, \ie, the sample of tangential
velocities from \Hipparcos\ used by \citet{1998AJ....115.2384D}, to
determine the velocity distribution and a different, non-overlapping,
sample of observations, \ie, radial velocity measurements of stars
from the \gcs, to validate the reconstructed velocity
distribution. Any structure in the velocity distribution that passes
this truly hard test can then confidently be considered real. We will
see that few do.

In addition to this, since we obtain a semi-parametric representation
of the three-dimensional velocity distribution we are equipped to
consider different subspaces of this distribution, such that we can
make predictions for the velocities of individual stars, both only
knowing their position as well as conditioning on any known components
of the velocities, which allows us to answer questions as to how much
information is contained in lower-dimensional projections of the
velocity distribution. For instance, we can ask how much information
about the velocity distribution is gained from a knowledge of the
radial velocities in addition to the tangential velocities. Such a
question is important to ask in the context of the \Gaia\ mission,
which will not measure radial velocities for a large part of its
catalogue \citep{2001A&A...369..339P}.

Finally, with the model of the velocity distribution in hand we can
determine the peaks in the distribution and compare them to the known
moving groups. The algorithm used in the optimization naturally
returns probabilities for individual stars to be members of moving
groups which allows us to study the members of the groups and, thus,
the properties of the moving groups themselves. This will be pursued
further in paper II in this series.

\section{Data, model, and algorithm}\label{sec:model}

Throughout we use the standard Galactic velocity coordinate system,
with the directions $x$, $y$, and $z$ (and associated unit vectors
$\eex$, $\eey$, and $\eez$) pointing toward the Galactic center, in
the direction of circular orbital motion, and toward the north
Galactic Pole, respectively. Vectors are everywhere taken to be column
vectors. The components of the velocity vector, $\eex\T\vv$,
$\eey\T\vv$, and $\eez\T\vv$, are conventially referred to as $U$,
$V$, and $W$, respectively, but we will refer to them as \vx, \vy, and
\vz.

\subsection{\Hipparcos\ measurements}

In this study we use a kinematically unbiased sample of 11,865 nearby
main-sequence stars \citep{1998MNRAS.298..387D} from the \Hipparcos\
catalogue \citep{ESA97a}. This sample is the union of two
kinematically unbiased subsamples of \Hipparcos\ stars: One subsample
consists of a magnitude-limited subsample (complete to about 7.3 to 9
mag depending on Galactic latitude) and the other subsample contains a
sample of stars south of $\dec =-28\degree$ judged from their spectral
classification to be within 80 pc from the Sun. Main-sequence stars
with relative parallax errors smaller than 10\,percent and not part of
a binary system were selected from both of these subsamples. The
relative parallax error cut, while biasing the sample towards brighter
stars, which have smaller parallax errors, as well as towards closer
stars, which have larger parallaxes, does not affect the fact that the
sample is kinematically unbiased, since the precision of the
parallaxes in \Hipparcos\ is mainly limited by Poisson noise and the
accuracy of the attitude reconstruction of the \Hipparcos\ satellite,
both of which are unrelated to the kinematics. For the stars in this
sample we take the equatorial coordinates, parallaxes, and proper
motions from the new reduction of the \Hipparcos\ data
\citep{2007ASSL..250.....V,2007A&A...474..653V}. Although follow-up
radial velocity measurements for the \Hipparcos\ stars were suggested
\citep{1989Msngr..56...12G}, radial velocities of many of the stars in
this sample are unavailable. Because of the known issues of including
radial velocities of only a subsample of stars
\citep[\eg,][]{1997ESASP.402..473B}, we give each star a radial
velocity of zero, with an uncertainty that is many orders of magnitude
larger than any of the velocities involved in this problem. For the
purposes of the deconvolution technique described below, this approach
is equivalent to an incomplete data approach, since the model and
objective function that we use properly treat the uncertainties
associated with the data (we have explicitly checked this).

\begin{figure}
\includegraphics[width=\textwidth]{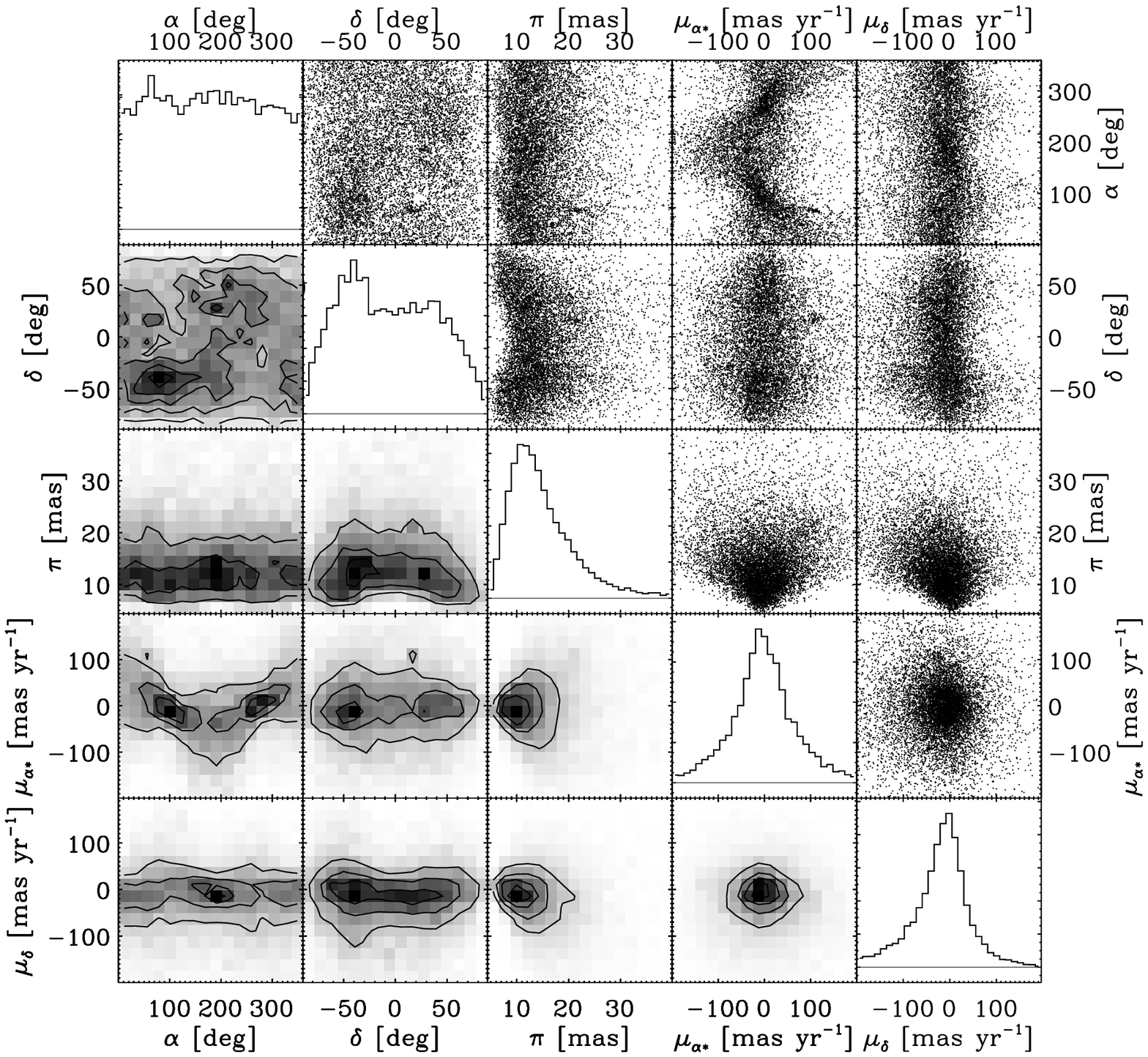}
\caption{Properties of the kinematically unbiased subsample of main-sequence stars extracted from the \Hipparcos\ catalogue. The diagonal shows histograms of the relative abundances of stars in the basic properties right ascension (\ra), declination (\dec), parallax ($\pi$), proper motion in \ra\ (\pmrastar), which includes a factor of $\cos \dec$, and proper motion in \dec\ (\pmdec). The plots in the upper-right triangle show two-dimensional scatter plots of pairs of these properties, while the lower-left triangle shows two-dimensional histograms for these pairs.}%
\label{fig:hip2prop}%
\end{figure}

The properties of the selected sample of stars in the observed
quantities are shown in \figurename~\ref{fig:hip2prop}. One can
clearly see the overdensities near the poles of the ecliptic in the
\ra\ vs.~\dec\ plots. These overdensities are a consequence of both
the scanning strategy of \Hipparcos, which covers stars near the
ecliptic poles much better than those near the ecliptic plane, leading
to higher accuracies of the \Hipparcos\ parallaxes near the ecliptic
poles (see also the parallax \parallax\ vs.~\dec\ plots), as well as,
for the south ecliptic pole, the inclusion of the sample of stars
restricted to $\dec \leq -28\degree$. The structure \pmra\ vs.~\ra\
and \pmra\ vs.~\dec\ panels is simply due to the Solar motion with
respect to the Local Standard of Rest.

The components of the three-dimensional velocities $\vv$ of the stars
in terms of the observed (\ra,\dec,\parallax,\pmra,\pmdec,\vrr) is
given by
\begin{equation}\label{eq:vrpmrapmdectoUVW2}
\vv \equiv \matrixleft \begin{array}{c} \vx \\ \vy \\ \vz \end{array} \matrixright =
\TT \, \AAA \, \matrixleft \begin{array}{c} \vrr  \\ \frac{k}{\parallax}\,\pmra\,\cos\dec\\\frac{k}{\parallax}\pmdec\end{array} \matrixright\, ,
\end{equation}
where $k =  4.74047$, [\vrr] = km s$^{-1}$, [\parallax] = \arcsecs,
[\pmra]=[\pmdec]= \arcsecs\ yr$^{-1}$, and the matrices $\TT$ and $\AAA$
are given by
\begin{equation}\label{eq:radectolbT}
\TT = \matrixleft \begin{array}{ccc} \cos \theta & \sin \theta & 0\\\sin \theta & -\cos \theta & 0\\0&0&1 \end{array} \matrixright
\matrixleft \begin{array}{ccc} -\sin \decngp & 0 & \cos \decngp\\ 0 & 1 & 0 \\ \cos \decngp & 0 & \sin \decngp\end{array} \matrixright
\matrixleft \begin{array}{ccc} \cos \rangp &\sin\rangp&0\\ -\sin\rangp & \cos\rangp&0\\ 0&0&1\end{array} \matrixright\, ,
\end{equation}
and
\begin{equation}
\AAA = \matrixleft \begin{array}{ccc} \cos \ra & -\sin \ra & 0 \\ \sin \ra & \cos \ra & 0 \\ 0 & 0& 1 \end{array} \matrixright
\matrixleft \begin{array}{ccc} \cos \dec & 0 & -\sin \dec\\ 0 & 1 & 0 \\ \sin \dec  & 0 & \cos\dec\end{array} \matrixright\, ,
\end{equation}
respectively. In the context of the deconvolution technique that we
use below to fit the velocity distribution we will define the
observations to be $\ww \equiv \matrixleft\begin{array}{ccc} \vrr &
\frac{k}{\parallax}\,\pmra\,\cos\dec&\frac{k}{\parallax}\pmdec\end{array}
\matrixright\T$ and the projection matrix to be $\RR^{-1} \equiv
\TT\,\AAA$. Since we do not use the radial velocities of the stars, we
set $\vrr$ to zero in $\ww$. The matrix $\TT$ depends on the epoch
that the data were taken at (1991.25 for \Hipparcos) through the
values of \rangp, \decngp, and $\theta$ (the position in celestial
coordinates of the north Galactic pole, and the Galactic longitude of
the north Celestial pole, respectively). These quantities were defined
for the epoch 1950.0 as follows: \citep{1960MNRAS.121..123B}: $\rangp
= 12^{\textnormal{h}}49^{\textnormal{m}}$, $\decngp = 27\degree.4$,
and $\theta = 123\degree$. This transformation is the only processing
of the data which we perform beyond the sample cut. This means that we
make no corrections for the effects of Galactic rotation---the effect
of which is of the order of the observational uncertainties
anyway---and do not subtract out the velocity of the local standard of
rest, since we simply want to study the distribution of stellar
velocities with respect to the Sun.

The \Hipparcos\ catalogue entries, which can be represented by some
vector $\cci$ for each star, come with single-star uncertainty
covariance matrices $\CCi$\footnote{These covariance matrices can be
constructed from the upper-triangular weight matrices $\UUi$ included
in the new reduction of the \Hipparcos\ data as
\begin{equation}
\CCi= \UUi^{-1}\,\left(\UUi^{-1}\right)\T\, .
\end{equation}}. If we write
the derivative of the observations $\wwi$ with respect to the
catalogue entries $\cci$ in terms of a matrix $\QQi$,
\begin{equation}
\dd \wwi = \QQi \dd \cci\, ,
\end{equation}
then the measurement uncertainty covariances $\SSi$ for the $\wwi$ are
given by
\begin{equation}
\SSi=\QQi \CCi \QQi\T\, .
\end{equation}
This is only accurate in the regime of small parallax errors in which
we are working. Star--to--star covariances in the \Hipparcos\ data
could be significant, \eg, through uncertainties in the modeling of
the sattelite altitude; they are believed to be insignificant and are
not reported in the \Hipparcos\ catalogue. Although significant
star--to--star correlations existed in the original \Hipparcos\
catalogue, these are reduced by a factor 30 to 40 in the new
reduction \citep[see \figurename~2.11 in][]{2007ASSL..250.....V}.

\subsection{Model and objective function}

The model for the velocity distribution and the objective function
which we use here has been described before
\citep{2005ApJ...629..268H} and is described in great detail in
\citet{Bovy09a}. Here we summarize the most important aspects of the
model and the objective function. We refer the reader to
\citet{Bovy09a} for a more detailed derivation of the following
results.

The method for density estimation from noisy data used here is
completely general, and we will describe it in general terms before
specifying it to the problem at hand. Our goal is to fit a model for
the distribution function of a $d$-dimensional quantity $\vv$ using
only a set of $N$ observational data points $\wwi$. In general, we
assume that these observations are noisy projections of the true
values $\vvi$
\begin{equation}
\wwi = \RRi \vvi + \mbox{noise}\, ,
\end{equation}
where the noise is drawn from a normal distribution with zero mean and
known covariance matrix $\SSi$. As described above, in this specific
application we use a formally infinite eigenvalue---in practice a
value much larger than any of the measured values $|\wwi|$---in the
covariance matrix for the missing radial velocity, in which case no
actual projection occurs and the operator $\RRi$ is simply a rotation
matrix, whose inverse is given by the product $\TT\AAAi$ as described
in the previous subsection. Although our method permits arbitrary
variances and covariances in the observed properties of any data point
(star), it assumes that there are no point--point (star--star)
covariances. This is only true approximately in most cases of
interest.

We will model the distribution function $p(\vv)$ of the true values
$\vv$ as a mixture of $K$ Gaussians:
\begin{equation}
p(\vv) = \sum_{j=1}^K \alphaj \normal(\vv|\mmj,\VVj)\, ,
\end{equation}
where the amplitudes $\alphaj$ sum to unity and the function
$\normal(\vv|\mm,\VV)$ is the $d$-dimensional Gaussian distribution
with mean $\mm$ and variance tensor $\VV$. We emphasize here that the
number of Gaussians $K$ is a free parameter describing qualitatively
different models for the distribution function and that its value
therefore needs to be set by a model comparison technique. We will
have much more to say about this later.

The probability of the observed data $\wwi$ given the model parameters
$\theta$ is then given by a simple convolution of the model with the
error distribution (or, in probabilistic language, by a
marginalization over the true values of the velocities)
\begin{equation}
p(\wwi|\theta) \equiv p(\wwi|\SSi,\RRi,\theta) =\sum_j\int_\vv \mathrm{d}\vv\, p(\wwi,\vv,j|\theta)\, ,
\end{equation}
which works out to be itself a mixture of Gaussians
\begin{equation}
p(\wwi|\theta)  = \sum_{j=1}^K \alphaj \normal(\wwi|\RRi\mmj,\TTij)\, ,
\end{equation}
where
\begin{equation}
\TTij = \RRi\VVj\RRi\T + \SSi\, 
\end{equation}
because of the convolution of the model Gaussians with the uncertainty
Gaussians..

For a given value of $K$, the free parameters of the density model can
then be chosen such as to maximize (the logarithm of) the total
probability of the data given the model, or equivalently, (the
logarithm of) the likelihood of the model given the data
\begin{equation}\label{eq:likelihood}
\phi = \sum_i \ln p(\wwi|\theta) = \sum_i \ln \sum_{j=1}^K \alphaj \normal(\wwi|\RRi\mmj,\TTij)\, .
\end{equation}
For each $K$, optimization of this function $\phi$ gives the best fit
density model consisting of $K$ Gaussians. This optimization could be
performed by any generic optimizer, such as nonlinear conjugate
gradients. It is complicated, however, by the constraints on the
parameters of the model, \eg, the amplitudes have to add up to one,
the covariances must be positive definite. For this reason we opt for
an optimization technique known as \emph{expectation--maximization}
\citep[\EM;][]{Dempster1977}, which views both the true values of the
velocities $\vvi$ as well as the components $j$ from which they were
drawn as hidden variables and optimizes the probability of the
full data. Readers not interested in this technique can safely skip
the next subsection.

\subsection{\EM\ optimization algorithm}

We will only give a brief, self-contained description of the \EM\
technique we used to optimize the likelihood of the model of the
velocity distribution function given a set of noisy, heterogeneous,
and incomplete observations. Part of this algorithm has been described
before \citep[see the appendix of][]{2005ApJ...629..268H} and a full
description and proof of the method outlined in this section can be
found in \citet{Bovy09a}. We would also like to point out that this
algorithm was developed independently before (Diebolt \& Celeux, 1989,
unpublished; Diebolt \& Celeux, 1990, unpublished) and applied to the
velocity distribution in the Solar neighborhood
\citep{1990A&A...236...95G,1997ESASP.402..519F} for a small number of
components $K$ and without any of the extensions described below.

The \EM\ algorithm works by introducing the following sets of hidden
variables: (1) for each observation $\wwi$ a set of ``indicator
variables'' $\qij$, which indicate for each component of the mixture
of Gaussians whether this velocity was drawn from it, (2) for each
observation $\wwi$ the true velocity $\vvi$. Because the velocities
are not actually drawn from single components but rather from the full
mixture, the indicator variables $\qij$ take values between zero and
one and correspond to the probability that a velocity $\wwi$ was drawn
from a component $j$. Given these hidden variables---equivalently,
given full data---the likelihood of the model is given by
\begin{equation}\label{eq:incompletefulllike}
\Phi = \sum_i \sum_j \qij \ln \alpha_j \normal(\vvi|\mmj,\VVj) \ .
\end{equation}
This likelihood function $\Phi$ can be optimized analytically. The
strategy of any \EM\ algorithm is now to take, in the first step, the
expectation of this full-data likelihood given the data and a previous
guess of the model parameters (this stage is appropriately called the
\emph{expectation step}) and then to maximize this expectation value
in the second step (the \emph{maximization step}). Given an initial
guess for the model parameters, these two steps are performed
iteratively until convergence, identified here, as is usual, by
extremely small incremental improvement in the logarithm of the
likelihood per iteration.

In taking the expectation of $\Phi$ we need the expectation of the
quantities $\qij$, $\vvi$, and $\vvi\vvi\T$ given the data and a
current guess for the model parameters. Using some standard results
from multivariate normal theory it is easy to show that
\begin{align}
\bbij &\equiv \langle\vvi|\wwi,\SSi,\RRi,\theta,j\rangle = \mmj + \VVj
\RRi\T\TTij^{-1} (\wwi - \RRi\mmj)\\
\BBij &\equiv \langle(\vvi-\langle\vvi\rangle)(\vvi-\langle\vvi\rangle)\T|\wwi,\SSi,\RRi,\theta,j\rangle
= \VVj - \VVj\RRi\T \TTij^{-1}\RRi \VVj \ ,
\end{align}
while the expectation of the indicator variables is simply given by
the posterior probability that $\wwi$ was drawn from component $j$
\begin{equation}\label{eq:qij}
\qij \leftarrow \langle\qij|\theta,\wwi,\SSi,\RRi\rangle = \frac{\alpha_j \normal(\wwi|\RRi\mmj,\TTij)}{\sum_k \alpha_k \normal(\wwi|\RRi\mmk,\TTik)}\, .
\end{equation}

Given the expectation of the full-data likelihood, it then follows
that the following update steps maximize this expectation value
\begin{eqnarray}\label{eq:Mstep}
\alphaj &\leftarrow& \frac{1}{N}\,\sum_i \qij\nonumber \\
   \mmj &\leftarrow& \frac{1}{\qqj}\,\sum_i \qij\,\bbij\nonumber \\
   \VVj &\leftarrow& \frac{1}{\qqj}\,\sum_i \qij
                     \left[(\mmj-\bbij)\,(\mmj-\bbij)\T + \BBij\right]\, ,
\end{eqnarray}
where $\qqj = \sum_i \qij$.  Using Jensen's inequality one can show
that these expectation and maximization steps also lead to a monotonic
increase in the probability of the observed data given the model
(given in \eqnname~[\ref{eq:likelihood}]).

Some problems that are commonly encountered using the \EM\ algorithm
to iteratively compute maximum likelihood estimates of the parameters
of a Gaussian mixture are singularities and local
maxima. Singularities arise when a Gaussian component becomes very
peaked or elongated. The standard method to deal with this problem is
to introduce a prior distribution for the model covariances, \eg, a
Wishart density \citep{Ormoneit1995}. Since the Wishart density is a
conjugate prior for the covariance of a multivariate normal
distribution \citep[\eg,][]{Gelman00a}, the update steps in
\eqnname~(\ref{eq:Mstep}) are modified only slightly (in the simplest
case) by the introduction of a regularization parameter $w$ in the
update step for the covariances
\begin{equation}\label{eq:Mstepreg}
   \VVj \leftarrow \frac{1}{\qqj+1} \,\left[\sum_i \qij
                     \left[(\mmj-\bij)\,(\mmj-\bij)\T+\BBij\right]+ w \II\right]\, ,
\end{equation}
where, again, $\qqj = \sum_i \qij$.  This regularization parameter is
another free parameter which is not known a priori, but should be
inferred from the data (in the context of Bayesian inference it is
known as a \emph{hyperparameter}). We will discuss its determination
in detail below.

The fact that the \EM\ algorithm monotonically increases the
likelihood of the model given the data is both one of the advantages
as well as one of the disadvantages of the \EM\ method. The algorithm
is very stable because of this property, leading to reasonable answers
largely irrespective of the initial guess for the model
parameters. However, because the likelihood cannot but increase in
every step, the algorithm can easily get stuck in a local maximum of
the likelihood function. In order to deal with this problem, we have
to discontinuously change the model parameters to jump to a different
region of parameter space. One way of doing this is by merging two of
the Gaussians in the mixture and splitting a third Gaussian, thus
conserving the number of Gaussians, after an initial run of the \EM\
algorithm, and reconverging \citep{Naonori1998}. The new solution is
then accepted if the likelihood of the model is larger than it was
before this \emph{split and merge} step. This procedure is halted
after a sufficiently large number of possible split and merge steps
fail to give an improvement of the likelihood of the model. We again
refer the reader to \citet{Bovy09a} for details on how the Gaussians
are merged and split and how the ranking of split and merge candidates
is established.

\subsection{The \gcs}\label{sec:gcs}

In what follows, we fit a model of the velocity distribution using
only the tangential velocities measured by \Hipparcos, and then later
we validate our results using radial velocity measurements. The radial
velocity measurements which we use for this purpose are all taken from
the \gcs\ \citep[\gcsabb;][]{2004A&A...418..989N}. The \gcsabb\
catalogue consists of metallicities, ages, and kinematics for a
complete, magnitude-limited, and kinematically unbiased sample of
16,682 nearby F and G dwarfs. We select from this sample all of the
stars that have a \Hipparcos\ entry and exclude any star that is
suspected to be a giant or part of a binary system; this leaves us
with 7,682 stars. For these stars we only take the radial velocities
and the uncertainty in the radial velocities from the \gcsabb\
catalogue, using the new reduction of the \Hipparcos\ data to provide
all of the other kinematical information. We performed no processing
of the data beyond the sample cut.

\section{Application to \Hipparcos\ data}

\begin{figure}
\includegraphics{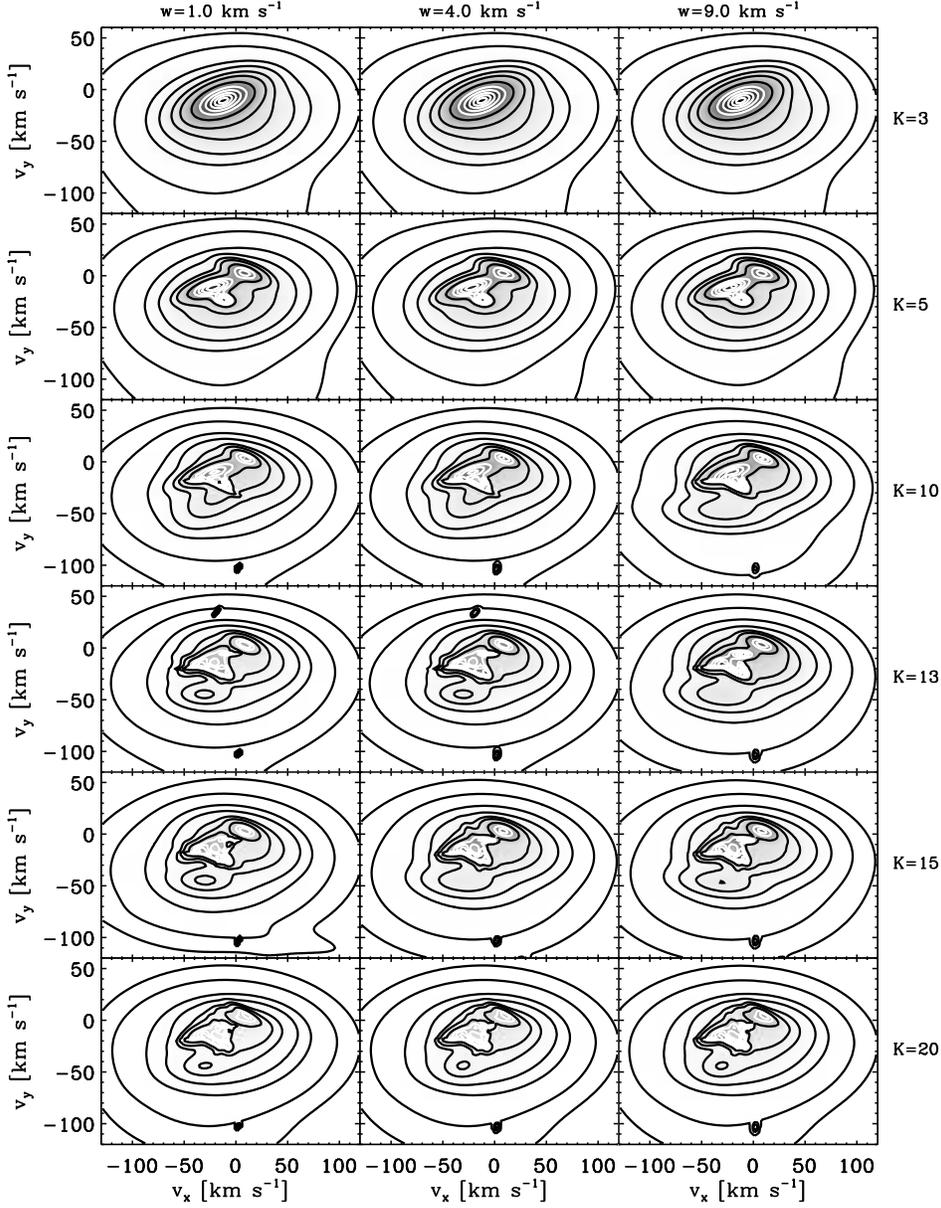}
\caption{Projection on the $v_x-v_y$ plane of the reconstructed velocity distribution as a function of the number of Gaussians $K$ and the regularization parameter $w$ used in the reconstruction. The density grayscale is linear and contours contain, from the inside outward, 2, 6, 12, 21, 33, 50, 68, 80, 90, 95, 99, and 99.9 percent of the distribution. The first five of these contours are white and somewhat blended together in some of the panels; 50 percent of the distribution is contained within the innermost dark contour. The origin in each of these plots is at the Solar velocity.}%
\label{fig:veldensXY}
\end{figure}

\begin{figure}
\includegraphics{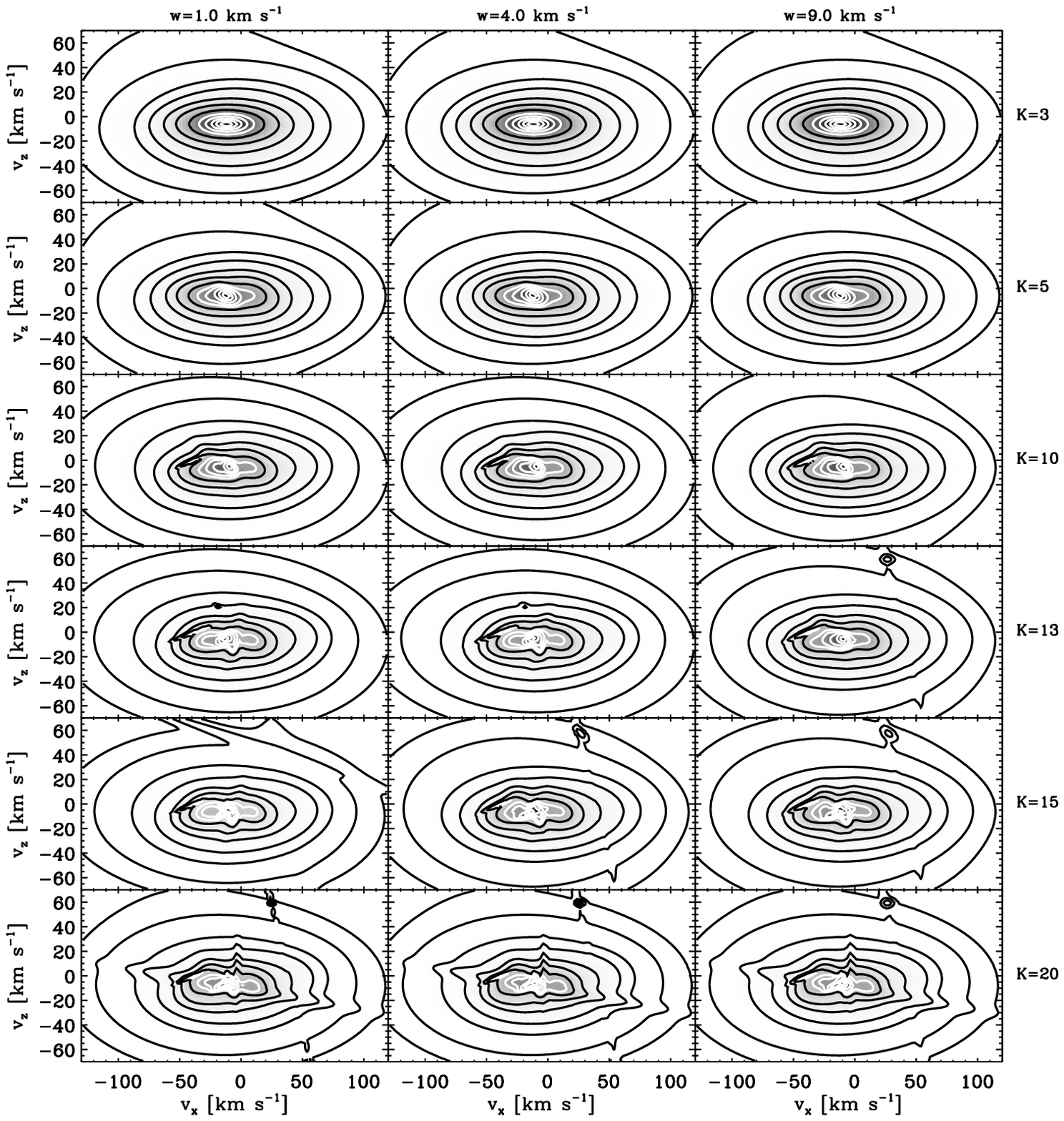}
\caption{Same as Fig.~\ref{fig:veldensXY}, but projected onto the $v_x-v_z$ plane.}%
\label{fig:veldensXZ}
\end{figure}

\begin{figure}
\includegraphics{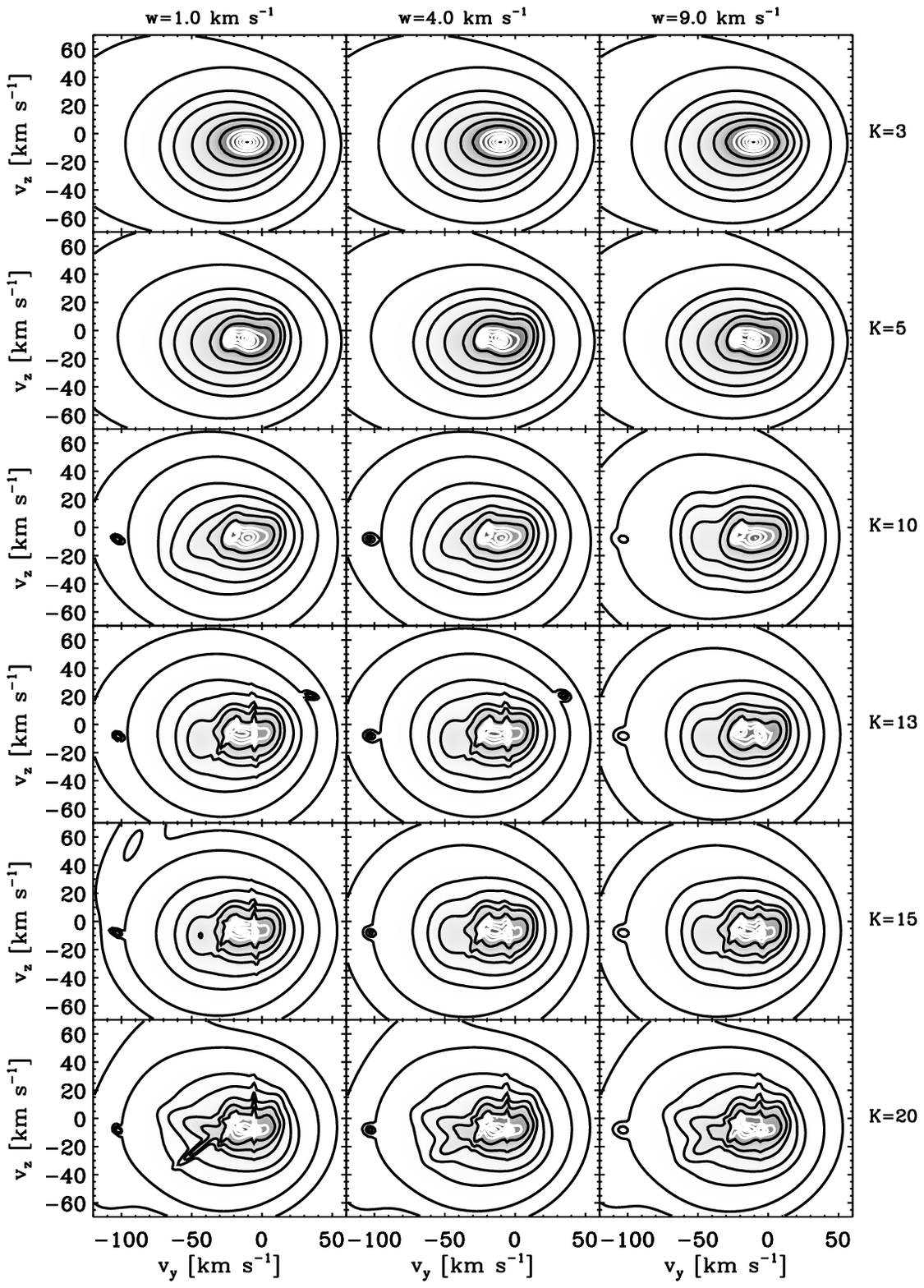}
\caption{Same as Fig.~\ref{fig:veldensXY}, but projected onto the $v_y-v_z$ plane.}%
\label{fig:veldensYZ}
\end{figure}

Two-dimensional projections of the reconstructed three-dimensional
velocity distribution for models characterized by different values of
the number of Gaussians $K$ and the regularization parameter $w$,
which, in a sense, sets the smallest scale on which we can infer
structure in the velocity distribution, are shown in
\figuresname~\ref{fig:veldensXY}-\ref{fig:veldensYZ}, although in high
density regions a large amount of data trumps the regularization as
can be seen from \eqnname~(\ref{eq:Mstepreg}).  The regularization
parameter $w$ is expected to be of order one because of the typical
magnitude of the velocity errors and because of the spatial range of
the sample of stars: a ``cold'' moving group on a circular orbit
spread out over the range of our sample, $\Delta x\approx 100$ pc,
will create elongated structures in the velocity distribution with a
size $\approx \Delta x/R_\odot \times 220$ km s$^{-1}$ $\approx$ 2.5
km s$^{-1}$; the typical uncertainties in the velocities are of the
same size and thus blur these elongated structures. 

For each ($K$,$w$) pair two runs of the optimization algorithm were
made. One started from randomly chosen initial conditions for all of
the components of the mixture, the other one started with the best
result obtained for the less complex models---models with lower $K$,
models with larger $w$, or both---and added a small-amplitude new
Gaussian component when necessary (this is unnecessary when the best
result for the less complex models occurs for a model with the same
$K$ but larger $w$). After optimization we then choose the
reconstructed velocity distribution with the highest likelihood.

The classical moving groups are present in each reconstruction of the
velocity field for all but the smallest values of $K$. Projections of
the distribution involving the $v_z$ component of the velocity are, as
expected, essentially featureless for all of the considered values of
$K$ and $w$. The similarity of all of the reconstructions shows that
the gross features---the rough overall shape and the main peaks---of
the velocity distribution are very robust to the model selection
question: Each of the models with $K \gtrsim 7$ reproduces the most
salient features of the velocity distribution.

\begin{figure}
\includegraphics{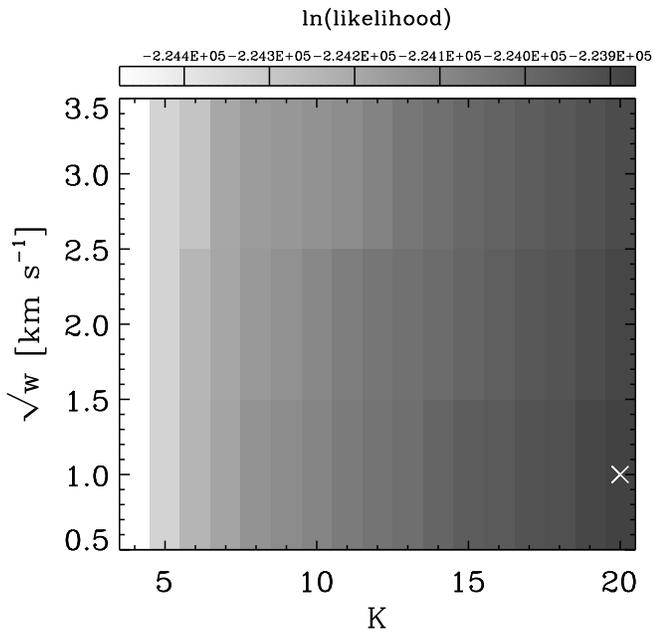}
\caption{Likelihood of the reconstructed velocity distribution given the tangential velocities of the \Hipparcos\ stars as a function of the number of Gaussians $K$ and the regularization parameter $w$ used in the reconstruction. The likelihood increases as the number of Gaussians is increased and as the regularization parameter is decreased. The white cross indicates the position of the maximum.}%
\label{fig:loglike}
\end{figure}

The logarithm of the likelihood of the different models given the
observed tangential velocities is shown in
\figurename~\ref{fig:loglike}. As complexity increases with increasing
$K$ (more components) and decreasing $w$ (less restrictions on the
covariances) the likelihood of the models increases in these
directions of increasing complexity; in
\figuresname~\ref{fig:veldensXY}-\ref{fig:veldensYZ} this increase in
complexity translates into smaller and smaller substructures coming to
the surface in the velocity distribution. The increasing complexity of
the velocity distribution is not confined to the projection onto the
Galactic plane of the velocity distribution, small-scale features can
also be seen in the projections that involve the vertical
direction. We could further increase the likelihood of our model of
the velocity distribution by adding more and more complexity to our
model. The more important issue here is how much complexity is
warranted by the data. This is the question which we discuss in the
next section.

\section{The complexity of the model}

In order to determine which combination of the model parameters $K$
and $w$ gives the best description of the underlying velocity
distribution we will perform a series of model selection tests. Most
of these tests are internal tests, meaning that they only use the data
that was used to fit the velocity distribution (in our case these are
the tangential velocities from the \Hipparcos\ data), but we will
perform one external test: We will use the reconstructed velocity
distributions for each set of $K$ and $w$ to predict the radial
velocities of the stars in the \gcsabb\ catalogue and test these
predictions. The preferred model is then the model that best predicts
the radial velocities. From the outset we can say that an external
test like this is to be preferred since it gives the strongest, most
independent test of the validity of the reconstructed model of the
velocity distribution.

\subsection{Internal model selection tests}

One of the simplest, and most robust, internal tests that we can
perform is leave-one-out cross validation \citep{stone74a}. In the
most ideal application of this technique one creates $N$ data samples
by leaving out one data point (the measurements for one star here) at
a time and performs the full maximum likelihood fit of the velocity
distribution for each of these data samples. Then one records the
logarithm of the probability of the data point that was left out given
the model found by fitting to all of the other data points and one
adds up all of the log probabilities found in this way. The model to
be preferred is then the model that gives the highest total
probability of the left out data points. Why will this work? This
procedure punishes models that overfit the data, that is, complex
models that use their complexity to fit very specific features
consisting of a small number of stars.

While simple, this technique can be computationally very expensive, as
the analysis, which can be hard for even one data set, has to be
repeated for $N$ data sets of roughly the same size, for each value of
the parameters $K$ and $w$. In practice one therefore often chooses to
leave out a certain fraction of the data sample at each stage, \eg,
1\,percent of the data, and record in each step the total probability
of the left out fraction of the data. However, for the current
application this still turns out to be much too computationally
expensive, and we therefore chose to only allow the amplitudes of the
component Gaussians to change from their maximum-likelihood values
from the global fit for each of the leave-one-out trials.

Akaike's information criterion (\AIC) is a model selection criterion
rooted in the concept of entropy, considering the amount of
information lost when representing the data by the model
\citep{Akaike}. We use an interpretation of the \AIC\ which was
developed for model selection in the context Gaussian mixture modeling
\citep{Bozdogan,Windham}. The \AIC\ is defined as
\begin{equation}
\mathrm{\AIC}(K,w)= -\frac{2}{N}\,\left[N-1-\nparam-\frac{K_{\mbox{{\footnotesize max}}}}{2}\right]\phi(K,w)+3\totalnparam\, ,
\end{equation}
where $K_{\mbox{{\footnotesize max}}}$ is the maximum number of components
one would consider (which we set to 100), \nparam\ is the number of
parameters per component, \totalnparam\ is the total number of
parameters estimated, and $\phi(K,w)$ is the logarithm of the
probability of the data given the best estimate of the velocity
distribution given $K$ and $w$.

Another set of model selection criteria make use of the principle of
minimum message length. According to this principle the best model of
the data is the model that allows for the shortest full description of
the data. It can be thought of as implementing Occam's razor in a more
sophisticated way than the chi--squared per degree of freedom
folklore. The message length corresponding to a given model consists
of the sum of the length of the message required to communicate the
model parameters and the length of the message that transmits the
residuals of the data given the model. As such, the message length is
equivalent to the Kolmogorov complexity, the length of the shortest
program that could output the data, which is, in general, incomputable
\citep{solomonoff64a,solomonoff64b,kolmogorov65a}. Therefore, in order
to create a working model selection criterion based on the principle
of minimum message length certain restrictions to the set of allowed
codes must be made \citep{Wallace99a}.

One such set of restrictions that does not make any assumptions about
the process that generated the data is offered by the minimum
description length principle \citep{rissanen78a,Grunwaldbook}. The
minimum description length is given by \citep{rissanen78a,schwarz78a}
\begin{equation}\label{eq:mdl}
\mathrm{\MDL}(K,w) = -\phi(K,w)+\frac{1}{2}\totalnparam\,\log N\, .
\end{equation}

Another, in some sense, Bayesian approach to the minimum coding inference
principle is given by minimum message length
\citep[\MML;][]{wallace68a,wallacebook}. In minimum message length
one's prior beliefs about the data generating process are used in full
in the encoding process, such that the message is formed by taking the
prior assumptions about the model together with the data to find the
shortest description of the data and the model. Minimum message length
goes through great pains to come up with the shortest message length,
leading to the following expression for the message length for the
case of Gaussian mixtures \citep{wallace87a,oliver94a,Oliver96a}
\begin{equation}\label{eq:mml}
\begin{split}
\mathrm{\MML}(K,w) = & K \log\left(2\,\det \VV_{\mbox{{\footnotesize pop}}}\right)
-\log(K-1)! +\frac{\totalnparam}{2}\log\kappa(\totalnparam) - \log K!\\
& + \sum_{i=1}^{d}\sum_{j=1}^K \log \frac{\sqrt{2}\,\alpha_j\,N}{\lambda_{j,i}}+
\frac{1}{2}\log N -
\frac{1}{2}\sum_{j=1}^K \log \alpha_j -
\phi(K,w) +\frac{\totalnparam}{2}\, .
\end{split}
\end{equation}
In this the $\lambda_{j,i}$ are the $d$ eigenvalues of $\VVj$ for each
component $j$; $\VV_{\mbox{{\footnotesize pop}}}$ is the observed
covariance matrix of the distribution. We determine this covariance
matrix $\VV_{\mbox{{\footnotesize pop}}}$ by fitting a single Gaussian
distribution to the observed sample of stars in the same way as we fit
multiple component mixtures to the velocity
distribution. $\kappa(\totalnparam)$ is the optimal lattice quantizing
constant in an $\totalnparam$-dimensional space. The reason this
optimal lattice quantizing constant appears in the \MML\ expression is
that in order to minimize the message length one has to find the
accuracy to which the model parameters are specified which minimizes
the message length, that is, one has to quantize the model parameter
space in an optimal way. This involves setting the overall accuracy
scale---by specifying the volume of a quantum---but also, for each
scale, finding the optimal arrangement of quantized values of the
model parameter space for that scale. This latter optimization amounts
to minimizing the squared-error made when quantizing and the optimal
lattice quantizing constant is the constant of proportionality between
the minimum squared-error and the product of the scale raised to the
appropriate power---$2/D$ when the scale is the quantum of volume in
the model parameter space---and $D$, the dimension of the space. For
example, optimal quantization of a one-dimensional quantity is
achieved by using intervals of constant width $s$ and the minimum
squared-error is given by $s^2/12$, the value of the squared error for
a uniform distribution. Therefore, the optimal lattice quantizing
constant in one dimension is equal to 1/12. The scale of quantization
itself is set based on the precision to which the model parameters are
known, which is approximated using the Fisher matrix in the expression
above.

In more than one dimension the optimal arrangement of quantized values
is not in general a simple cubic lattice---indeed, this optimal
arrangement is unknown in more than three dimensions---and the value of
the optimal lattice quantizing constants are not known in more than
three dimensions either \citep{conway92a}, although tight bounds on
the value of the optimal lattice quantizing constant exist
\citep{zador63,zador82}
\begin{equation}\label{eq:zador}
\frac{1}{(n+2)\pi} \Gamma\left[\frac{n}{2}+1\right]^{2/n} \leq
\kappa(n) \leq \frac{1}{n\pi} \Gamma\left[\frac{n}{2}+1\right]^{2/n}\,
\Gamma\left[1+\frac{2}{n}\right]\, ;
\end{equation}
as $n$ goes to infinity, $\kappa(n) \rightarrow 1/2\pi e$. For our
purposes the dimensionality at which we evaluate these lattice
quantizing constants are very high and the results we find later do
not depend on whether we choose the upper or the lower bound in
\eqnnumber~(\ref{eq:zador}). The expression for the shortest message
length given in \eqnnumber~(\ref{eq:mml}) is still only an approximate
expression, assuming that the components of the mixture do not overlap
significantly. This is obviously not the case in the application to
the velocity distribution, as can be seen from
\figuresname~\ref{fig:veldensXY}-\ref{fig:veldensYZ}, but it is a
necessary assumption in order to calculate the shortest message
length.

Finally, we also use the actual Bayesian model selection criterion,
known as the \emph{evidence}. The evidence is defined as the
denominator in the application of Bayes's theorem to parameter
estimation, which in this case given by
\begin{equation}\label{eq:bayes}
p(\theta | \mbox{\Hipparcos\ data}, \{K, w\}) = 
\frac{p(\mbox{\Hipparcos\ data} | \theta)\,
p(\theta|\{K,w\})}{p(\mbox{\Hipparcos\ data}|\{K,w\})}\, ,
\end{equation}
in which $p(\mbox{\Hipparcos\ data} | \theta)$ is the likelihood of
the model. Choosing uninformative priors for the model parameters, the
evidence can be approximated as
\begin{equation}
\begin{split}
\log p(\mbox{\Hipparcos\ data}|\{K,w\}) = &
\phi(K,w) - K\, \log\left(2\,\det \VV_{\mbox{{\footnotesize pop}}}\right)
+\log \left(K-1\right)!\\ & +\frac{\totalnparam}{2}\log (2\pi)
-\frac{1}{2} \left(\sum_{j=1}^{K-1} \log\sum_{i=1}^N 
\left(\frac{\qij}{\alpha_j}-\frac{q_{iK}}{\alpha_K}\right)^2\right. \\
&\left.+2\, d\sum_{j=1}^K\log \left(\sqrt{2} \alpha_j N\right)
-2 \sum_{j=1}^K \sum_{i=1}^d \log \lambda_{j,i}\right)
\end{split}
\end{equation}
in the case of a Gaussian mixture \citep{roberts98a}. The $\qij$ are
defined in \eqnnumber~(\ref{eq:qij}), and the other quantities
appearing in this equation have the same meaning as in
\eqnnumber~(\ref{eq:mml}).

Next we will discuss an external test for the complexity of the
reconstructed velocity distribution.

\subsection{Validation with the \gcsabb\ radial velocities}

Given our three-dimensional reconstruction of the velocity
distribution in the Solar neighborhood, based solely on the tangential
velocities from \Hipparcos, we can validate each model of the velocity
distribution using an external data set. The external data set which
we use here is a set of radial velocities from the \gcs, which was
described above in \sectionname~\ref{sec:gcs}. For each set of $K$ and
$w$ we proceed as follows: for each star in the sample taken from the
\gcsabb\ we predict the distribution of the radial velocity from the
reconstructed velocity distribution and we record the probability of
the actual measured radial velocity given this predicted probability
distribution for the radial velocity. The ``best'' model for the true
velocity distribution is then given by that set of ($K$, $w$) that
leads to the overall highest probability of the measured radial
velocities.

How do we predict the probability distribution of the radial velocity
for a given star given our three-dimensional model for the velocity
distribution?  We can distinguish two distinct predictions: we can
base our prediction on the position of the star on the sky, or we can
base our prediction both on the position of the star as well as on the
observed tangential velocity of the star. It is this last prediction
which we will use in order to compare different models for the
velocity distribution.

Given the position of a star on the sky we know the radial direction
$\eer$ from which we can construct the projection onto the radial
direction. We define the radial and tangential projection operators by
\begin{equation}
\radialproj \equiv \eer\eer\T \qquad \qquad \tangproj \equiv \unitmatrix - \eer\eer\T\, ,
\end{equation}
in which $\unitmatrix$ is the unit matrix. We can then decompose the
velocity distribution as
\begin{equation}\label{eq:radialdecomp}
p(\vv) = \sum_{j=1}^K \alpha_j \normal(\vv|\radialproj \mmj+
\tangproj \mmj, \radialproj \VVj\radialproj +\tangproj \VVj\tangproj + \radialproj\VVj\tangproj + \tangproj\VVj\radialproj )\, ,
\end{equation}
Marginalizing over the value of the tangential velocity is then simply
performed by dropping the tangential directions from these Gaussians,
such that
\begin{equation}
p(v_r|\ra,\dec) = \sum_{j=1}^K \alpha_j\,
\normal(v_r|\radialproj \mmj,\radialproj \VVj\radialproj)\, .
\end{equation}
In order to obtain the probability of an observed radial velocity
given this prediction, we need to go one step further. Since the
measured radial velocities come with their own uncertainties, we need
to convolve this predicted distribution with the error distribution of
the measured value. Assuming that this error distribution is a
Gaussian with zero mean and variance $\sigma_r^2$, the predicted
distribution becomes
\begin{equation}\label{eq:margpredicted}
p(v_r|\ra,\dec) = \sum_{j=1}^K \alpha_j\,
\normal(v_r|\radialproj \mmj,\radialproj \VVj\radialproj +\sigma_r^2)\, .
\end{equation}

Since we know the tangential velocities of the stars in the \gcsabb,
we can use this information to make a more informed prediction for
that star's radial velocity. In order to do this we start again from
\eqnnumber~(\ref{eq:radialdecomp}), but this time we condition this
distribution on the value of the tangential velocity. For each of the
Gaussian components in \eqnnumber~(\ref{eq:radialdecomp}) this
conditioning basically comes down to performing a linear regression
along the principal axes of the Gaussian ellipsoids, evaluated at the
value of the tangential velocity. This linear regression needs to be
performed while taking into account the uncertainties in all of the
quantities, \ie, the full error covariance matrix of the tangential
velocities $\SSt$, and the uncertainty on the radial velocity
$\sigma_r^2$ (since the uncertainties in the radial velocities are
uncorrelated with the \Hipparcos\ uncertainties). Using some standard
results from multivariate normal theory the predicted distribution of
the radial velocity of a star $i$ given its position and tangential
velocity follows:
\begin{equation}\label{eq:condpredicted}
p(v_r|\ra,\dec,\pmra,\pmdec,\parallax) = \sum_{j=1}^K \qij\,
\normal(v_r|\mmrj,\TTrj)\, ,
\end{equation}
in which $\qij$ is calculated as in \eqnnumber~(\ref{eq:qij}),
\begin{align}
\mmrj&\equiv\radialproj \mmj + \radialproj \VVj\tangproj  
\left( \tangproj \VVj\tangproj  + \SSt\right)^{-1}
\left(\vv_t-\tangproj \mmj\right)\, ,\\
\TTrj&\equiv\radialproj \VVj\radialproj +\sigma_r^2- \radialproj \VVj\tangproj 
\left( \tangproj \VVj\tangproj  + \SSt\right)^{-1} 
\tangproj \VVj\radialproj
\, ,
\end{align}
and $\vv_t$ is the tangential velocity.

In \figurename~\ref{fig:predict_gcs_rand} the marginalized and
conditioned predictions as well as the observed value are shown for a
random sample of radial velocities from the \gcsabb, for the
particular values of $K$ and $w$ that we will adopt below as our
fiducial values. The marginalized predictions are simply slices
through the velocity distribution in the radial direction for that
star, and therefore they are all rather broad. The conditioned
distribution in many cases shifts much of the mass of the probability
distribution to one or two sharp peaks, as the tangental velocities
pick out the most probable clumps that the star could be a part of.

The predicted distribution of the radial velocity of a particular,
random star from the \gcsabb\ as a function of the model parameters
$K$ and $w$ is shown in \figurename~\ref{fig:tile_onepred}. This shows
how the predictions of our models change as we increase the complexity
of the model. Only very subtle differences between the predictions can
be seen in this figure, as even the least complex models perform well
on this star. We do see the distribution tightening around the
observed value. In the model with the most complexity the extra
structure in the velocity distribution is not supported by this
particular star. By making these predictions as a function of $K$ and
$w$ we can answer the question of how much of the complexity of the
velocity distribution is warranted by the observed data.

\subsection{Assessing the complexity of the velocity distribution}\label{sec:complexity}

\begin{figure}
\includegraphics[width=.315\textwidth]{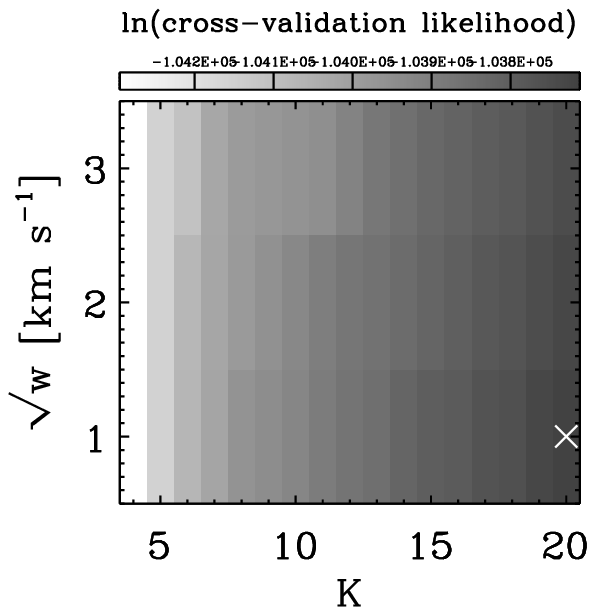}\hfill
\includegraphics[width=.315\textwidth]{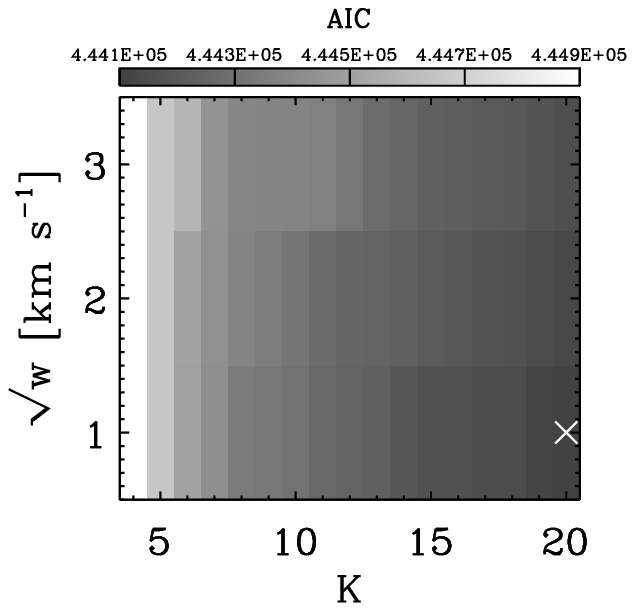}\hfill
\includegraphics[width=.315\textwidth]{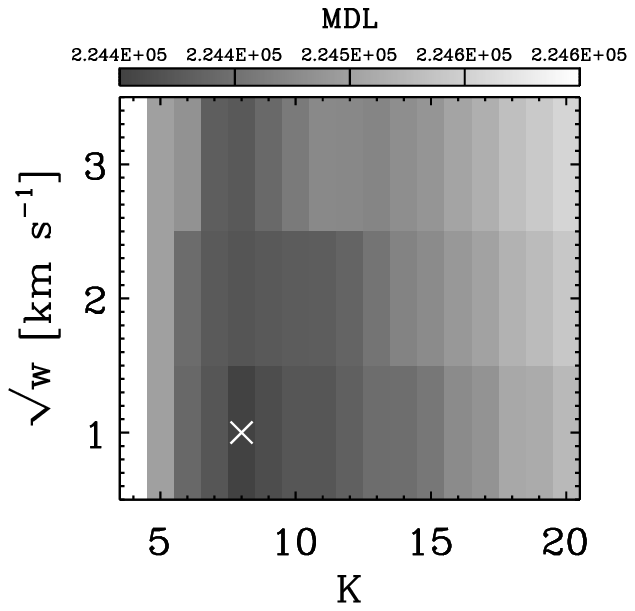}\\[10pt]
\includegraphics[width=.315\textwidth]{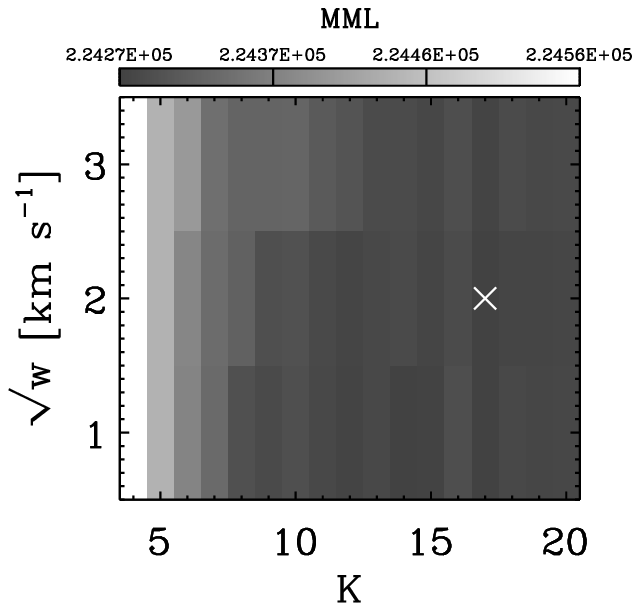}\hfill
\includegraphics[width=.315\textwidth]{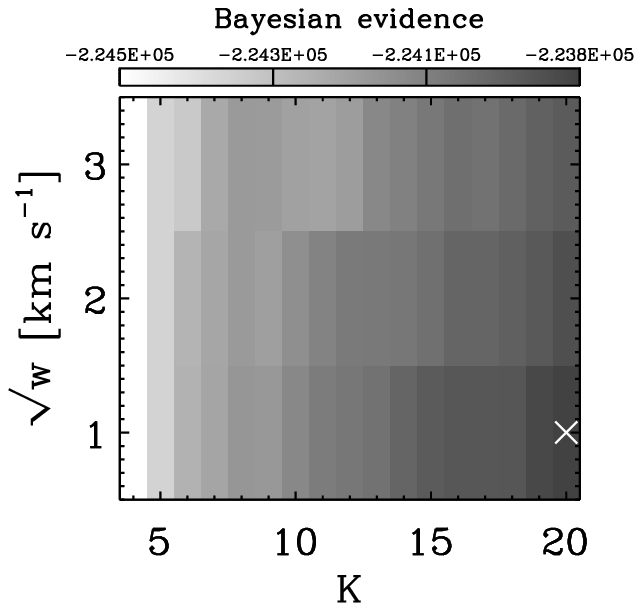}\hfill
\includegraphics[width=.315\textwidth]{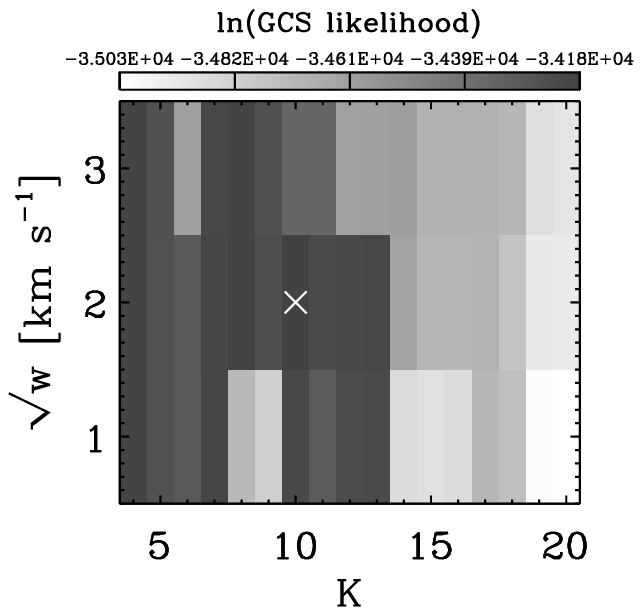}
\caption{Model selection surfaces: These surfaces show the different model selection criteria defined in the text applied to the reconstruction of the velocity distribution from \Hipparcos\ data. Models are defined by the number of Gaussian components $K$ and a regularization parameter $w$. In each of these figures a darker color represents a model that is more favored by the model selection criterium at hand; the white cross indicates the most favored model for each model selection criterium. Shown are (from left to right and from top to bottom): (1) cross-validation; (2) Akaike's Information Criterion (\AIC); (3) Minimum Description Length (\MDL); (4) Minimum Message Length (\MML); (5) Bayesian evidence; and (6) the likelihood of the predicted radial velocity distribution using radial velocities from the \gcsabb\ catalogue.}%
\label{fig:modelselection}
\end{figure}

Now that we have introduced the different model selection tests we can
apply them to our reconstructed models of the velocity distribution
and decide which combination of $K$ and $w$ is the most suited to
describe the velocity distribution. The model selection surfaces for
all of the different tests described above are given in
\figurename~\ref{fig:modelselection}. In each of the panels the darker
values of the density map correspond to models that are preferred by
that particular model selection criterion. We see that most of the
internal model selection criteria don't give a definite answer as to
the amount of complexity that is warranted by the data, as they prefer
models of ever increasing complexity. Only the \MDL\ criterion, which,
as can be seen from \eqnname~(\ref{eq:mdl}), is very harsh on the
introduction of extra parameters, seems to turn over around a value of
$K = 10$.

The one test that does have a clear preference is the test based on
the radial velocities from the \gcsabb. This model selection criterion
clearly prefers models with a value for the regularization parameter
$w$ of 4 km$^2$ s$^{-2}$, and among those models it prefers a moderate
value of $K$ = 10. Adding more and more components to the mixture
makes the predictions of the radial velocities progressively
worse. Since this test with the external radial velocities is arguably
the most stringent, we will adopt from now on the values $K = 10$ and
$w = 4$ km$^{2}$ s$^{-2}$ as fiducial values. The parameters of the
best fit model with these parameters are given in
Table~\ref{table:param}. We will discuss the features of this
reconstruction of the velocity distribution extensively below.

\begin{figure}
\includegraphics{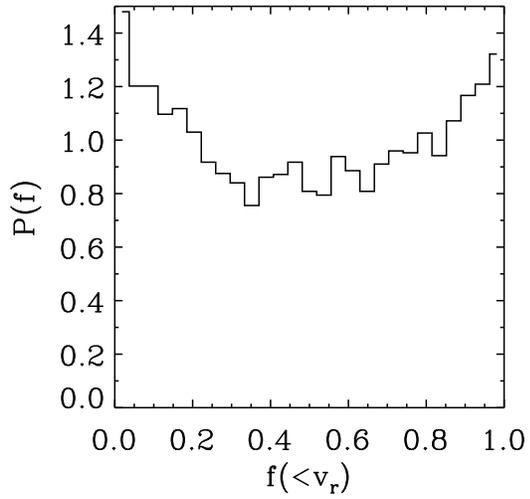}
\caption{Distribution of the quantile of the predicted radial velocity distribution (using $K$ = 10 and $w$= 4 km$^2$ s$^{-2}$) at which the radial velocity from the \gcsabb\ catalogue is found for all the stars from the sample we selected from the \gcsabb\ catalogue. If the probability distribution of the radial velocity for each star was entirely correct this curve should be flat at $P(f) = 1$.}%
\label{fig:checkquants}
\end{figure}

That our fiducial model does a good job of predicting the radial
velocities---\ie, it does not just do a better job than the other
models---can be seen from \figurename~\ref{fig:checkquants}, in which
the distribution of the quantiles of the predicted radial velocity
distribution at which the observed value of the radial velocity is
found is shown: For each radial velocity from the \gcsabb\ we
integrated the predicted radial velocity distribution up to the
observed value of the radial velocity and plotted the distribution of
the quantiles thus obtained. If our predicted distribution function
for the radial velocities was absolutely perfect this curve would be
completely flat. The fact that it rises slightly at the ends of the
interval is because our velocity distribution does a bad job of
predicting the radial velocities of high-velocity stars. However, for
intermediate and low velocity stars our predicted radial velocities
are in good agreement with the observations.

In the next section we discuss in more detail how well our
reconstruction of the velocity distribution predicts the radial
velocities of the \gcsabb\ stars, which will allow us to assess the
usefullness of complementary radial velocity surveys to upcoming
astrometric surveys for the purpose of establishing the statistical
properties of the kinematics of the stars.

\section{Information content of the predicted radial velocities}\label{sec:infcont}

\begin{figure}
\includegraphics{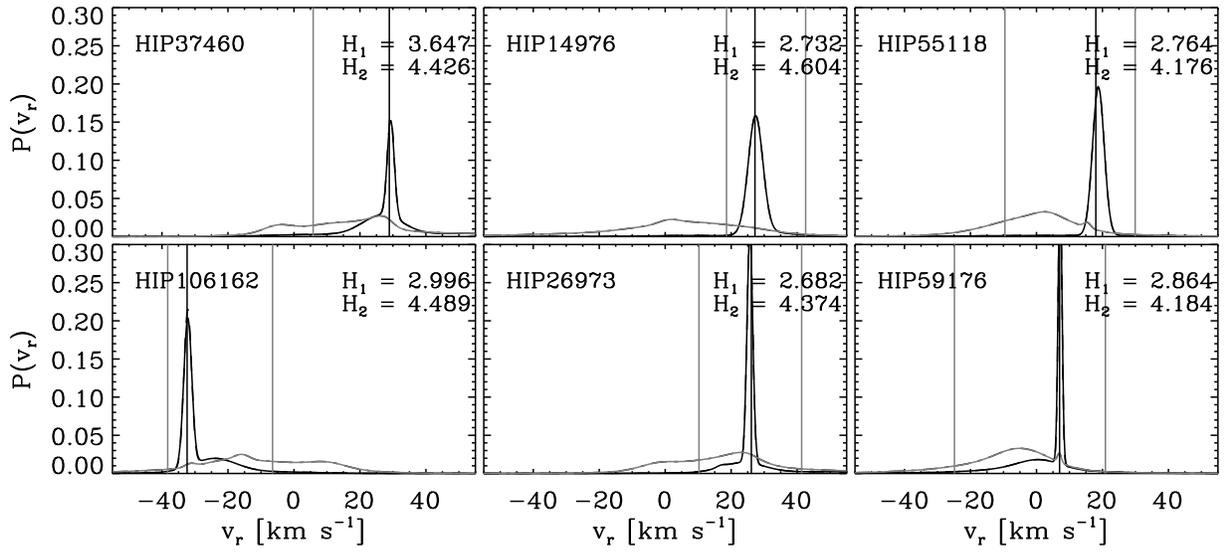}
\caption{The six ``best'', \ie, highest likelihood, predictions of the radial velocity of stars in the \gcsabb\ catalogue based on our reconstruction of the velocity distribution with $K$ = 10 and $w$= 4 km$^2$ s$^{-2}$.}%
\label{fig:predict_gcs_good}
\end{figure}

To start off the discussion on how well the reconstructed velocity
distribution predicts the radial velocities of stars in the \gcsabb\
we draw attention to some of the best predictions in
\figurename~\ref{fig:predict_gcs_good}. Shown are the six ``best''
predictions of radial velocities, where by best we mean that the
probabilities of the observed radial velocities given the model and
their tangential velocities are very large for these stars. What we
see are very narrow predicted distributions, with a width of only a
few km s$^{-1}$, and the observed radial velocities are right at the
peak of the distribution. The predicted distributions of these radial
velocities given only their position, which are shown in gray, are
very broad, such that the tangential velocities really pin down the
value of the radial velocity, as can also be seen by the large
difference in entropy between these two distributions. Thus, in these
cases the radial velocity of the star does not provide much extra
information, as its location in velocity space can be constrained
tightly from its tangential velocity alone.

\begin{figure}
\includegraphics{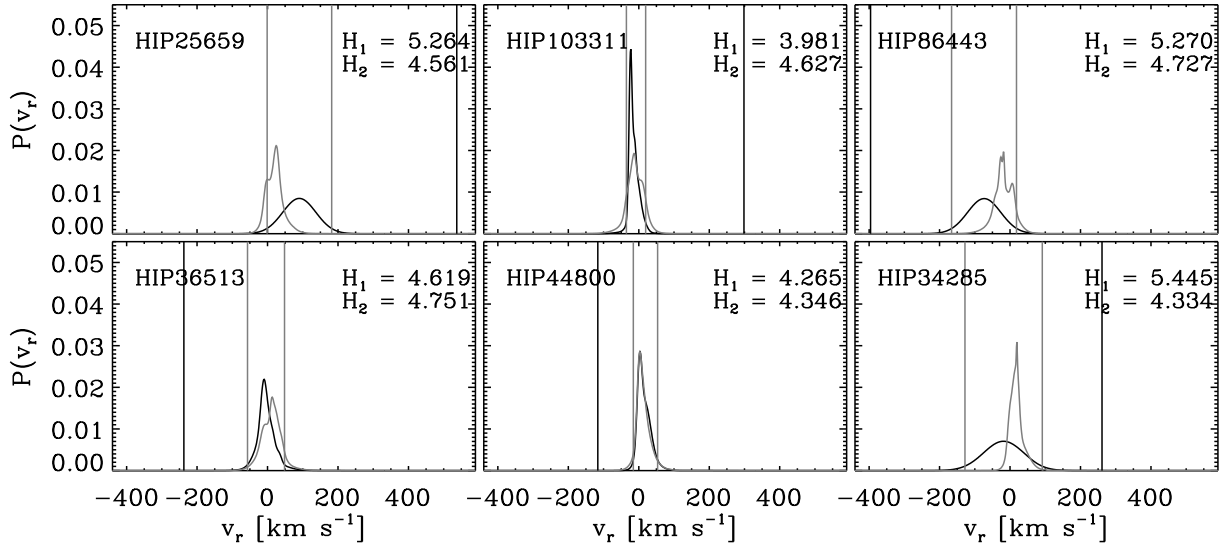}
\caption{Same as Fig.~\ref{fig:predict_gcs_good}, but the six ``worst'', \ie, lowest likelihood, predictions.}%
\label{fig:predict_gcs_bad}
\end{figure}

However, among the individual predictions we make there are also some
impressive failures. The six ``worst'' predictions are shown in
\figurename~\ref{fig:predict_gcs_bad}, where worst means here that
these radial velocities have very small probabilities given the model
and their tangential velocities. All of these bad predictions are for
very high velocity stars and it is no surprise that we cannot
accurately predict the radial velocities of these stars, since the
sample of stars that we used to reconstruct the velocity distribution
does not sample the high velocity stars, \eg, halo stars,
well. Keeping in mind the range in radial velocity in these plots, one
can see that the predicted distributions are very broad, with a
typical width of 100 km s$^{-1}$, and that the addition of the
tangential velocities to the positions of these stars in order to make
the prediction does not help to zero in on the value of the radial
velocity. Indeed, in most cases the entropy of the two predicted
distributions are about the same, and in a few cases the entropy of
the probability distribution for the radial velocity given the
tangential velocity is actually larger than the entropy of the
predicted distribution given the position of the star alone,
indicating that our model of the velocity distribution really has no
clue about the value of the radial velocity for these stars.

\begin{figure}
\includegraphics{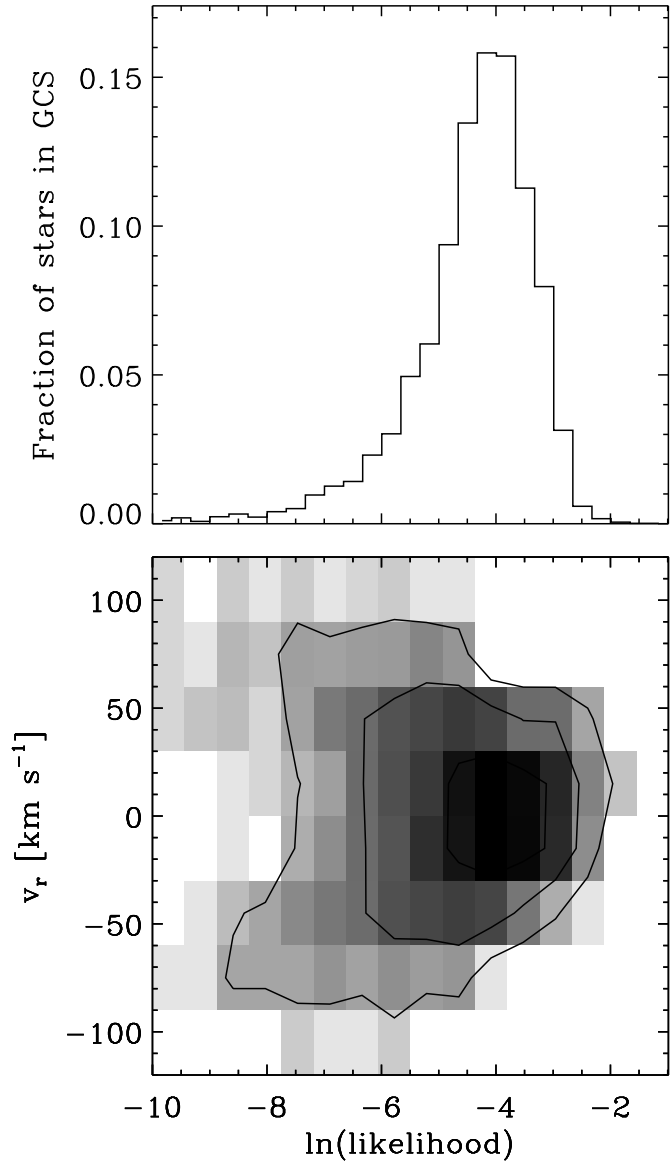}
\caption{Top: distribution of the likelihood of the predicted radial velocity distribution (with $K$ = 10 and $w$ = 4 km$^2$ s$^{-2}$) given stars from the \gcsabb\ catalogue. Bottom: two-dimensional histogram of the radial velocities from the \gcsabb\ catalogue and their probability under the reconstructed velocity distribution (gray scales are logarithmical).}%
\label{fig:hist_gcs_like}
\end{figure}

In \figurename~\ref{fig:hist_gcs_like} we look at the probability of
all of the radial velocities given the model and their tangential
velocities for the full sample of radial velocities from the
\gcsabb. A long tail towards small probabilities stands out in this
figure. The stars with radial velocities with the lowest probability
making up this tail are distributed all over the sky such that this
tail does not indicate that our deconvolution technique has failed to
reconstruct the velocity distribution in a particular direction (or
set of directions) on the sky. Inspecting the correlation between the
probability of the observed radial velocity given the model and the
value of the radial velocity shows that the tail is a consequence of
our inability to predict the radial velocities of very high velocity
stars, as we already established above. Ignoring these high-velocity
stars, we see that on average the probability of an observed radial
velocity given the model of the velocity distribution is $\approx
0.02$. In the language of information theory, this means that an
observed radial velocity adds about 4 nats $\approx$ 5.8 bits of
information to our knowledge.

\begin{figure}
\includegraphics{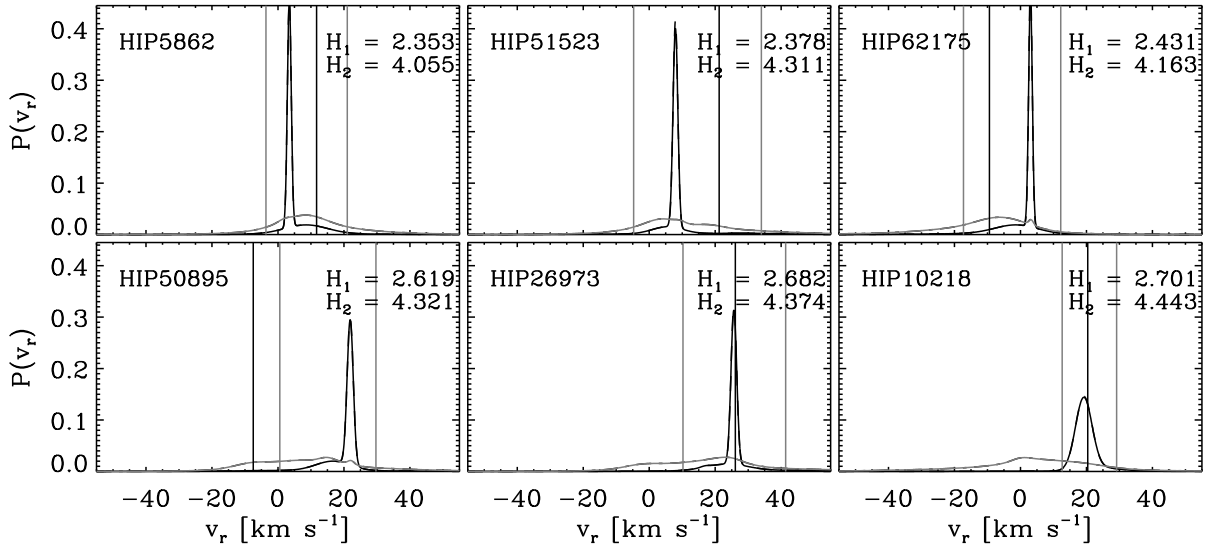}
\caption{The six ``tightest'', \ie, lowest entropy, predictions of the radial velocity of stars in the \gcsabb\ catalogue based on our reconstruction of the velocity distribution with $K$ = 10 and $w$= 4 km$^2$ s$^{-2}$.}%
\label{fig:predict_gcs_low}
\end{figure}

Another important measure of the information contained in our
predicted probability distributions for the radial velocities is the
entropy of the distribution. Broad probability distributions have
large entropies and low information content: They do not make a very
definite prediction for the radial velocity of a star. Very narrow,
sharp distribution functions have low entropy and do make tight
predictions for the radial velocities. In
\figurename~\ref{fig:predict_gcs_low} the six ``tightest'', or lowest
entropy, predicted probability distributions for the radial velocities
are shown. The probability distributions given the tangential
velocities are all very sharp, with typical widths of approximately 5
km s$^{-1}$ for these tightest predictions. The entropies of these
predictions based on the tangential velocities of the stars are all
much smaller than the entropies of the radial velocity distributions
based on the position alone, which again means that the tangential
velocities together with the model of the velocity distribution
strongly constrain the value of the radial velocity of these
stars. For this sample of the six tightest predictions, the predicted
radial velocities do not agree particularly well with the observed
radial velocities. In many cases the observed radial velocity is
located in a region of the probability distribution that is quite far
removed from the peak of the distribution, although in most cases
there is still a non-vanishing mass of the probability distribution
associated with the region of the observed radial velocity. This
clearly indicates that a knowledge of the tangential velocities alone
is not enough to reliably predict the radial velocity of a star, and
that the tangential velocity of a star can be very misleading in this
respect.

\begin{figure}
\includegraphics{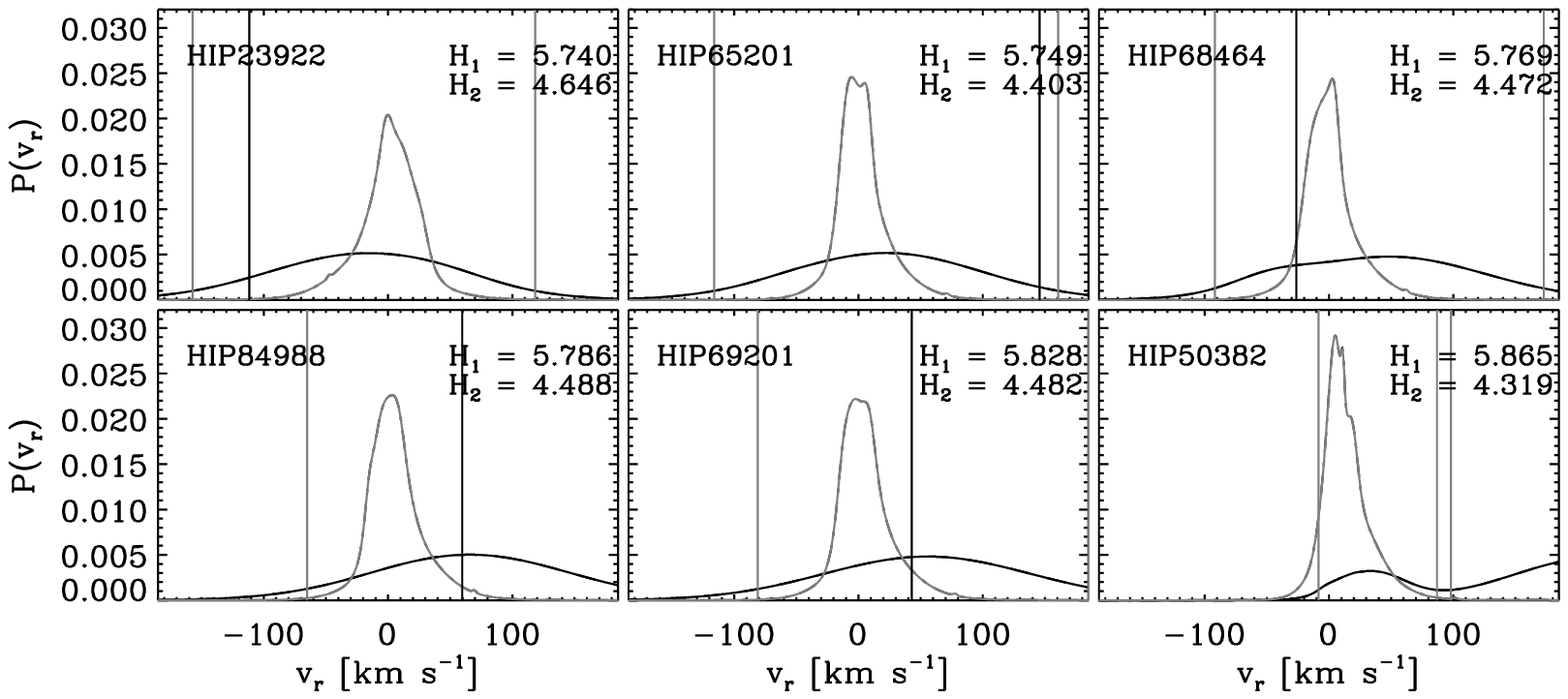}
\caption{Same as Fig.~\ref{fig:predict_gcs_low}, but the six ``widest'', \ie, highest entropy, predictions.}%
\label{fig:predict_gcs_high}
\end{figure}

The six ``widest'', or highest entropy, predicted probability
distributions are shown in \figurename~\ref{fig:predict_gcs_high}. The
predictions for the radial velocitites of these stars based on the
position alone are all informative, so these stars are such that the
direction in which they are observed has significant substructure that
could potentially pinpoint the velocity of the star based on the
tangential velocities of the star alone, but the measured tangential
velocity of the star does not indicate that the star is part of any of
the clumps in this direction. Most of the observed radial velocities
indeed do not correspond to any of the peaks in the predicted
distribution in the direction of the star. The probabilities of the
radial velocities of these stars given the model and their tangential
velocity is not particularly low compared to the average probability
of a radial velocity. Therefore, we can conclude that most of these
stars are probably part of the background population of stars,
although some, but not all, are also high-velocity stars.

\begin{figure}
\includegraphics{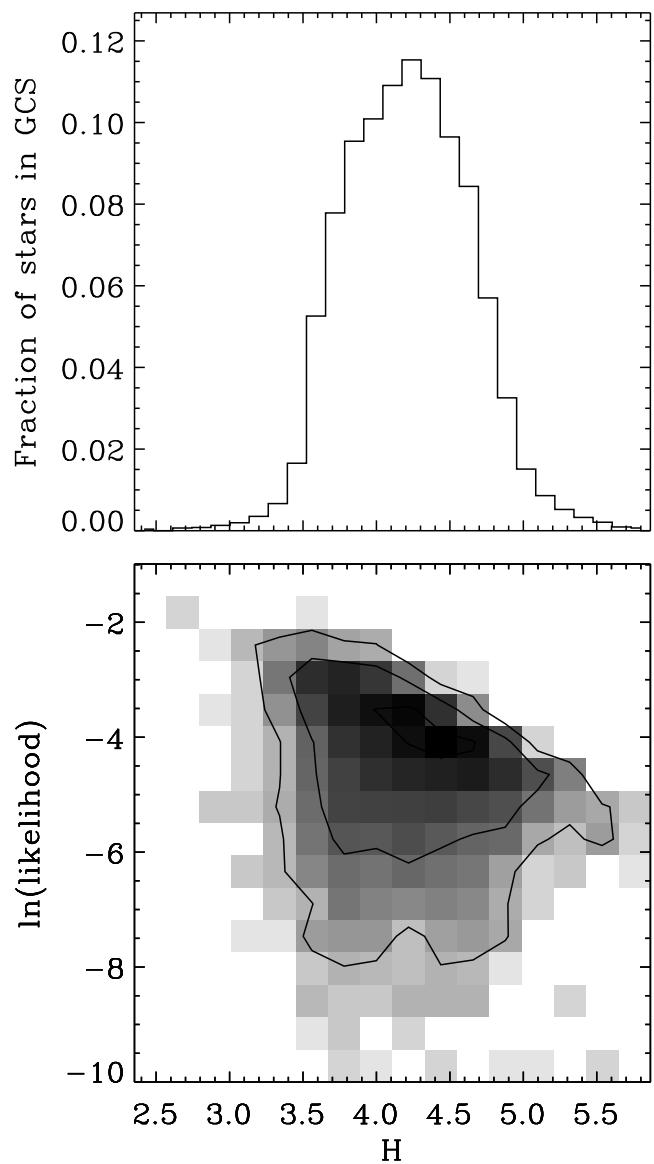}
\caption{Top: distribution of the entropy of the predicted radial velocity distribution (with $K$ = 10 and $w$ = 4 km$^2$ s$^{-2}$) for stars from the \gcsabb\ catalogue. Bottom: two-dimensional histogram of the likelihood of the predicted radial velocity distributions given stars from the \gcsabb\ catalogue and the entropy of the predicted distribution (gray scales are logarithmical).}%
\label{fig:hist_gcs_ent}
\end{figure}

In \figurename~\ref{fig:hist_gcs_ent} the distribution of the
entropies of the predicted probability distribution for the radial
velocities based on their tangential velocities is shown together with
a two-dimensional histogram of the entropies and the probabilities of
the observed radial velocities. The distribution of entropies is
fairly symmetrical around the mean value corresponding to a moderately
informative distribution. The bottom panel shows a hint of an
anti-correlation between the entropy of the predicted distribution and
the probability of the observed radial velocity: More informative
predicted distributions have a slight preference toward higher
probabilities of the observed radial velocity given this
distribution. This indicates that when the model together with the
observed tangential velocity makes an informative prediction,
corresponding to a low entropy, this prediction more often than not
turns out to have been a good prediction. However, the broad swath of
average entropies with low probabilities of the observed radial
velocity again indicates that our predicted radial velocities lack the
accuracy to make the observations of radial velocities obsolete in
this context.

\begin{figure}
\includegraphics{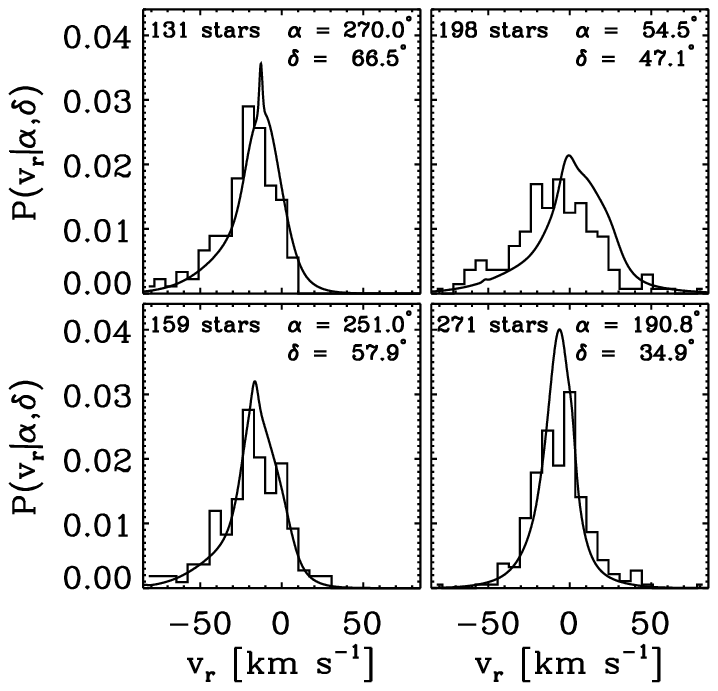}
\caption{Predicted radial velocity distribution for stars in different directions on the celestial sphere and observed distribution from the \gcsabb\ catalogue for different (\ra,\dec)-patches on the sky. Stars are selected to lie within 20$\degree$ of the central \ra\ and within 10$\degree$ of the central \dec, or in the corresponding region around the opposite \ra\ and \dec. The predicted radial velocity distribution is calculated by marginalizing the reconstructed velocity distribution using $K$= 10 and $w$ = 4 km$^2$ s$^{-2}$ at the center of each patch. The central \ra\ and \dec\ are given in the upper-right corner of each panel. The number of stars in the \gcsabb\ sample in the relevant (\ra,\dec)-patch are given in the upper-left corner of each panel.}%
\label{fig:radecpatches}
\end{figure}

As a final comparison of our predictions for the distributions of the
radial velocities we look at the distribution of radial velocities in
different patches of the sky. In \figurename~\ref{fig:radecpatches}
the predicted and observed distribution of radial velocities is shown
for four different directions: the direction towards the poles of the
ecliptic and three random directions. The observed distribution of
radial velocities is obtained by binning the radial velocities of
stars from the \gcsabb. The predicted distribution of radial
velocities is the prediction based on the central \ra\ and \dec\ of
each patch, as given in \eqnnumber~(\ref{eq:margpredicted}). The
predicted and observed distributions agree well for all of these
patches, as one can readily see by eye. Slight offsets between the
observed and predicted distributions are inevitable because of the
large range in both \ra\ and \dec\ that we have to give to the patches
in order to get a reasonable number of observed stars in a patch: the
distribution in a patch is non-uniform and this moves the observed
distribution away from the predicted distribution in some
cases. However, even in such cases, the shape of the predicted and
observed distributions agrees well. We refrain from making any
quantitive statements about the agreement between the observed and
theoretical distribution, \eg, by using the Kolmogorov--Smirnov
statistic for testing whether observations were drawn from a given
distribution, because of these difficulties in the interpretation of
the agreement between observed and predicted distribution. However,
this comparison does show that our reconstructed velocity field gets
the major properties of the distribution of the radial velocities
right. Thus, we see that we can predict the bulk properties of the
radial velocities from observations of the tangential velocities
alone. We stress the importance of the full sky-coverage of our sample
of tangential velocities in this respect: Under the assumption of a
homogeneous velocity distribution in this volume around the Sun we
need at least $2\pi$ sterradian to sample all directions of the
velocity distribution from the tangential velocities alone.

\section{Predictions for non-\gcsabb\ stars}

\begin{figure}
\includegraphics{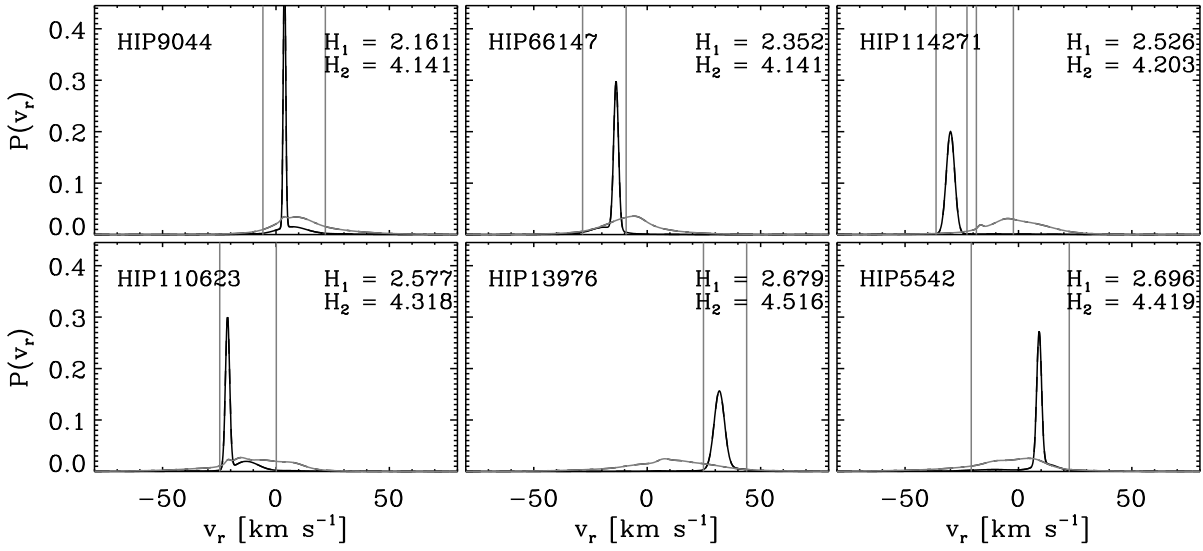}
\caption{The six ``tightest'', \ie, lowest entropy, predictions of the radial velocity of stars in the sample we extracted from the \Hipparcos\ catalogue that do not have an entry in the \gcsabb\ catalogue based on our reconstruction of the velocity distribution with $K$ = 10 and $w$= 4 km$^2$ s$^{-2}$.}%
\label{fig:info_hip_low}
\end{figure}

Since we can make detailed predictions for the radial velocity of any
star in the Solar neighborhood, whether we know its tangential
velocity or not, we can look at the stars from our sample from
\Hipparcos\ which do not have a measured radial velocity, and identify
particularly interesting stars from the point of view of their radial
velocity. For instance, we can calculate the entropy of the predicted
distribution for the radial velocity of a star and find the stars for
which our prediction is particularly informative. In
\figurename~\ref{fig:info_hip_low} the six most informative
predictions, \ie, lowest entropy predicted probability distributions
for the radial velocity, are shown for stars in our \Hipparcos\ sample
without radial velocity in the \gcsabb. The predicted distribution for
the radial velocity of these stars are all highly peaked at one
particular value for the radial velocity.

A detailed list of the properties of the stars with the most
informative predicted radial velocity distributions is given in
Table~\ref{table:hip_low}. This table lists the stars sorted by the
value of the entropy of their predicted probability distributions for
the radial velocity and gives the maximum likelihood estimate of the
radial velocity as well as 95\,percent upper and lower limits on the
radial velocity derived from the predicted distribution. As discussed
in the previous section and shown in
\figurename~\ref{fig:predict_gcs_low}, these predictions could still be
far fom the truth. 

\begin{figure}
\includegraphics{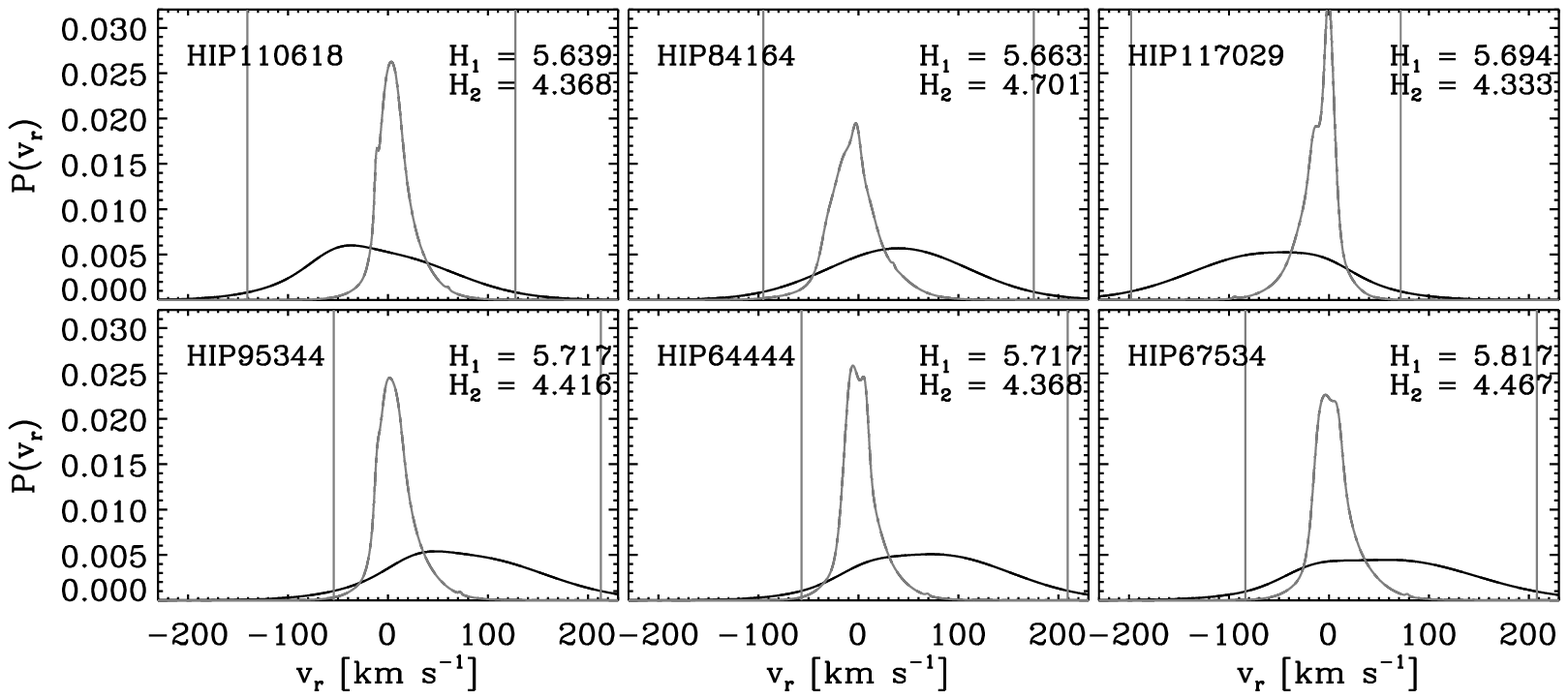}
\caption{Same as Fig.~\ref{fig:info_hip_low}, but the six ``widest'', \ie, highest entropy, predictions.}%
\label{fig:info_hip_high}
\end{figure}

A similarly interesting sample of stars are the stars for which the
predicted distribution of the radial velocity is very
uninformative. The six least informative predictions are shown in
\figurename~\ref{fig:info_hip_high}. As these are stars for which we
have essentially no clue about their radial velocity, given the model
of the velocity distribution and their tangential velocity, obtaining
their radial velocity would be interesting as it would add a
substantial amount of knowledge about the velocities of nearby
stars. A list of the 25 least informative predictions for stars in the
\Hipparcos\ sample used here that have no radial velocity from the
\gcsabb\ is given in Table~\ref{table:hip_high}.

\begin{figure}
\includegraphics{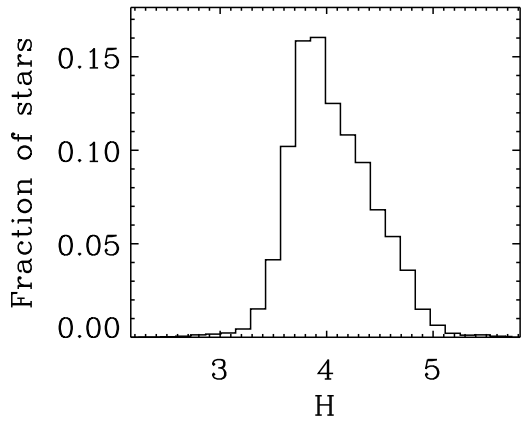}
\caption{Distribution of the entropy of the predicted radial velocity distribution (with $K$ = 10 and $w$ = 4 km$^2$ s$^{-2}$) for stars in the sample we extracted from the \Hipparcos\ catalogue that do not have an entry in the \gcsabb\ catalogue.}%
\label{fig:hist_hip_ent}
\end{figure}

Finally, \figurename~\ref{fig:hist_hip_ent} shows the distribution of
the entropies of the predicted distributions for all of the stars for
which we have no radial velocity. Comparing this distribution to the
distribution of the entropies of the predicted radial velocities of
stars which do have a measured radial velocity, shown in
\figurename~\ref{fig:hist_gcs_ent}, shows that the information content
and its distribution in this sample of unobserved radial velocities is
about the same as the information content in the stars that already
have measured radial velocities.

\section{Discussion}

\subsection{Why do the different model selection criteria differ?}

The analysis of the complexity of the velocity distribution in
\sectionname~\ref{sec:complexity} showed that there is a large
difference in this context between the internal model selection
criteria and the external test based on the predictions of our model
of the radial velocities of the \gcsabb\ stars. Because of the
importance of this conclusion for the comparison of our results with
the amount of structure in the velocity distribution reported in the
literature, we would like to understand the discrepancy between the
conclusions of internal and external tests of the structure in the
velocity distribution.

The first thing to note in this respect is that the specific forms of
the internal tests which we have applied are in most cases
approximations to the underlying model selection principles. For
example, in order to obtain a computationally feasible version of the
leave-one-out cross-validation principle we were forced to perform a
very restricted fit of the model to each of the leave-one-out
subsamples, \ie, we only allowed the best-fit model for each of the
leave-one-out subsamples to differ from the global best fit model in
the amplitudes of the Gaussian components of the mixture. Similarly,
the expression in \eqnnumber~(\ref{eq:mml}) for the message length
used in the minimum message length model selection criterion as well
as the form given in \eqnnumber~(\ref{eq:bayes}) for the Bayesian
evidence are only approximations to the true message length and the
true Bayesian evidence, respectively, for the Gaussian mixture model
for a distribution function. In the case of the approximation to the
message length we know that one of the approximations we are making is
wrong on some level, since the approximation assumes that the
different Gaussian components in the mixture do not overlap
significantly, whereas it is obvious from the reconstructions of the
velocity distribution in \figurename~\ref{fig:veldensXY} that this is
not the case. It is therefore possible that some of these
approximations are inappropriate for the current application and that,
if this were possible, the application of the true model selection
principle would give different results for the amount of complexity in
the velocity distribution. However, the fact that most of the internal
model selection criteria in \figurename~\ref{fig:modelselection}
behave similarly as a function of the complexity of the model hints at
the existence of a deeper reason for the discrepancy between the
internal and external model selection tests.

One of the reasons for this discrepancy could be unknown issues with
the data. For example, it is possible that significant star--to--star
covariances exist in the \Hipparcos\ data. Because of the complexity
of the \Hipparcos\ mission it is not immediately obvious that such
star--to--star covariances should be small, although much effort has
been made in the data reduction process to remove these correlations
\citep[\eg,][]{2007ASSL..250.....V}. Unknown correlations between
supposedly independent data points can cause internal model selection
criteria such as leave-one-out cross validation to overestimate the
amount of structure in the velocity distribution function because the
``external'' check of the velocity distribution found for each
jackknife-subsample with the data point that was left out fails, since
if the data point that was left out is correlated with other data
points it was not really left out of the data set. Other internal
model selection criteria which we used above will be similarly
affected, as, for instance, unknown star--to--star covariances will
lead to an overestimate of the message length in the minimum message
length criterion which could easily move the minimum of the message
length function towards higher complexities.

Another possible issue with the data is a difference between the
assumed, Gaussian error distribution for the \Hipparcos\ parallaxes
and proper motions and the true error distribution, either because of
underestimates of the width of the error distribution or through
non-Gaussianities. Both of these possibilities would generically lead
to a a larger amount of structure in the velocity distribution than
the actual amount of structure. 

Rather than issues with the data there could be problems with the
model for the velocity distribution which we use here. If the model is
inadequate, \ie, if the true velocity distribution does not lie in or
even near the space of allowed models, then model selection will fail
in general, because one is choosing between different models which are
all bad representations of the true velocity field. In that situation
it is natural to find that the model selection tests prefer models of
increasing complexity, because, \eg, adding a component could
substantially increase the goodness of the fit over the model with
less complexity, while still being very far from the truth. However,
it seems unlikely that the true velocity distribution is very far
removed from being adequately described by a mixture of Gaussian
distributions model, since the velocity distribution of a relaxed
population of stars is approximately Gaussian and dispersing star
clusters should also be well-described by approximately Gaussian
velocity distributions. However, if the structure in the velocity
distribution is caused by resonances, or if the projection onto
velocity space of partially phase-mixed structures in the
six-dimensional phase-space distribution of stars gives rise to
singularities \citep{1999MNRAS.307..877T}, such as folds or cusps,
then the mixture of Gaussian distributions model might be inadequate.

Although it is hard to assess the role played by the different effects
described in this subsection in determining the outcome of the model
selection tests, it should be clear that an external model selection
tests will be affected much more weakly by all of them. The validation
of the reconstruction of the velocity distribution by the radial
velocities of \gcsabb\ stars does not require us to make any
approximations, since we can predict straightforwardly the probability
distribution for the radial velocity of each \gcsabb\ star and compare
it to the observed radial velocity. The external test is not plagued
by any unknown issues with the \Hipparcos\ data since the radial
velocity data is completely separate from the \Hipparcos\ data (as we
project our model of the velocity distribution onto the space of the
radial velocity for each \gcsabb\ star rather than combining the
\gcsabb\ radial velocity with the tangential velocities measured by
\Hipparcos\ to form the three-dimensional velocity $\vv$). The
external test also provides a robust check of the adequacy of the
model space which we consider, as the radial velocities would all be
very improbable if the model were very far from the truth. We saw that
this was not the case in Section~\ref{sec:infcont}. The external test
applied here is therefore the most conservative model selection
test. Any structure in the velocity distribution that passes this
conservative test can therefore be reliably considered real. This does
not, however, exclude the existence of more structure in the velocity
distribution, although if more structure exists, it must be at much
lower significance in the \Hipparcos/\gcsabb\ data than the structure
recovered here.

\subsection{The complexity of the velocity distribution}

As we discussed in \sectionname~\ref{sec:complexity}, our model
selection criterion based on the \gcsabb\ radial velocities shows a
clear preference for a model of the velocity distribution with only a
modest amount of complexity. The preferred model consists of ten
Gaussian components such that only a limited number of clumps are
apparent in the reconstructed velocity distribution, which is shown in
\figurename~\ref{fig:annotated_veldist}. This stands in sharp contrast
with many of the analyses of the velocity distribution in recent
years, which have revealed an increasing amount of structure and
clumps in the velocity distribution.

One of the first comprehensive studies of the velocity distribution
based on \Hipparcos\ data showed lots of structures in the
$\vx$--$\vy$ projection of the velocity distribution
\citep{1998AJ....115.2384D}. Subsequent analyses turned up even more
substructures, in the form of branches \citep{1999MNRAS.308..731S} and
an ever-increasing number of clumps, or moving groups candidates
\citep{2008A&A...490..135A,2009ApJ...692L.113Z}. Using much of the
same data we are unable to confirm much of the perceived structure. A
few reasons are responsible for this: (1) our rigorous, conservative
model selection procedure and (2) the attention we paid to the
smallest scales on which we can reliably detect substructures given
our data.

Comparing the criteria which we used to decide on the best model of
the velocity distribution among many models characterized by different
levels of complexity shows that internal tests of the complexity of
the velocity distribution, such as the criteria that have been applied
in all of the recent studies, give very different answers than model
selection criteria depending on an external data set (see also the
discussion in the previous subsection). Much of the evidence for the
existence of structure in the velocity distribution in the past has
hinged upon the detection of similar structures in different
subsamples of the data, \eg, color subsamples
\citep{1998AJ....115.2384D}, or on the detection of the same features
across different analysis techniques. All of these different
procedures for establishing the reality of the complexity seen in the
data all point towards a maximal amount of structure in the velocity
distribution, to such an extent that every little wiggle
\citep{1998AJ....115.2384D} or overdensity in the velocity
distribution \citep{2009ApJ...692L.113Z} is perceived as real. The
internal model selection criteria which we have applied here also lead
to the same conclusion: most of them prefer the models with the most
complexity. However, the external model selection criterion which we
applied, predicting the radial velocities of a large sample of stars
and preferring the model which best predicts these radial velocities,
is arguably more stringent than any internal test could ever be and
clearly gives a very different answer than the internal tests. This
test of predicting the radial velocities and comparing them to the
observations unambiguously points toward a model with only a modest
amount of complexity. 

A second, somewhat less important, reason for the discrepancy between
our results and some recent findings is that we explicitly considered
the smallest scales on which we could reliably detect structure in the
distribution of velocities, through our use of a renormalization
parameter $w$ for the covariances of the components of the
distribution, which roughly sets the scale of the smallest structures
we could find. We applied our model selection criteria to find the
best value of this parameter as well and found a clear preference for
a value of $w$ $\approx$ 4 km$^2$ s$^{-2}$ over smaller values. Thus,
testing our models of the velocity distribution on the radial
velocities from the \gcsabb\ shows that on average we cannot trust
structures on scales smaller than $\approx$ 2 km s$^{-1}$. Thus it is
unsurprising that we cannot recover structures in the velocity
distribution found by techniques such as wavelet analysis which
manually or semi-automatically set the scale of the structures they
want to find to values of 1 km s$^{-1}$ and smaller
\citep{2008A&A...490..135A,2009ApJ...692L.113Z}. As has been remarked
before \citep{1998AJ....115.2384D} and as our analysis also shows,
structures on scales of a few km s$^{-1}$ are likely to be spurious.

\subsection{The structure of the velocity distribution}

\begin{figure}
\includegraphics[width=\textwidth]{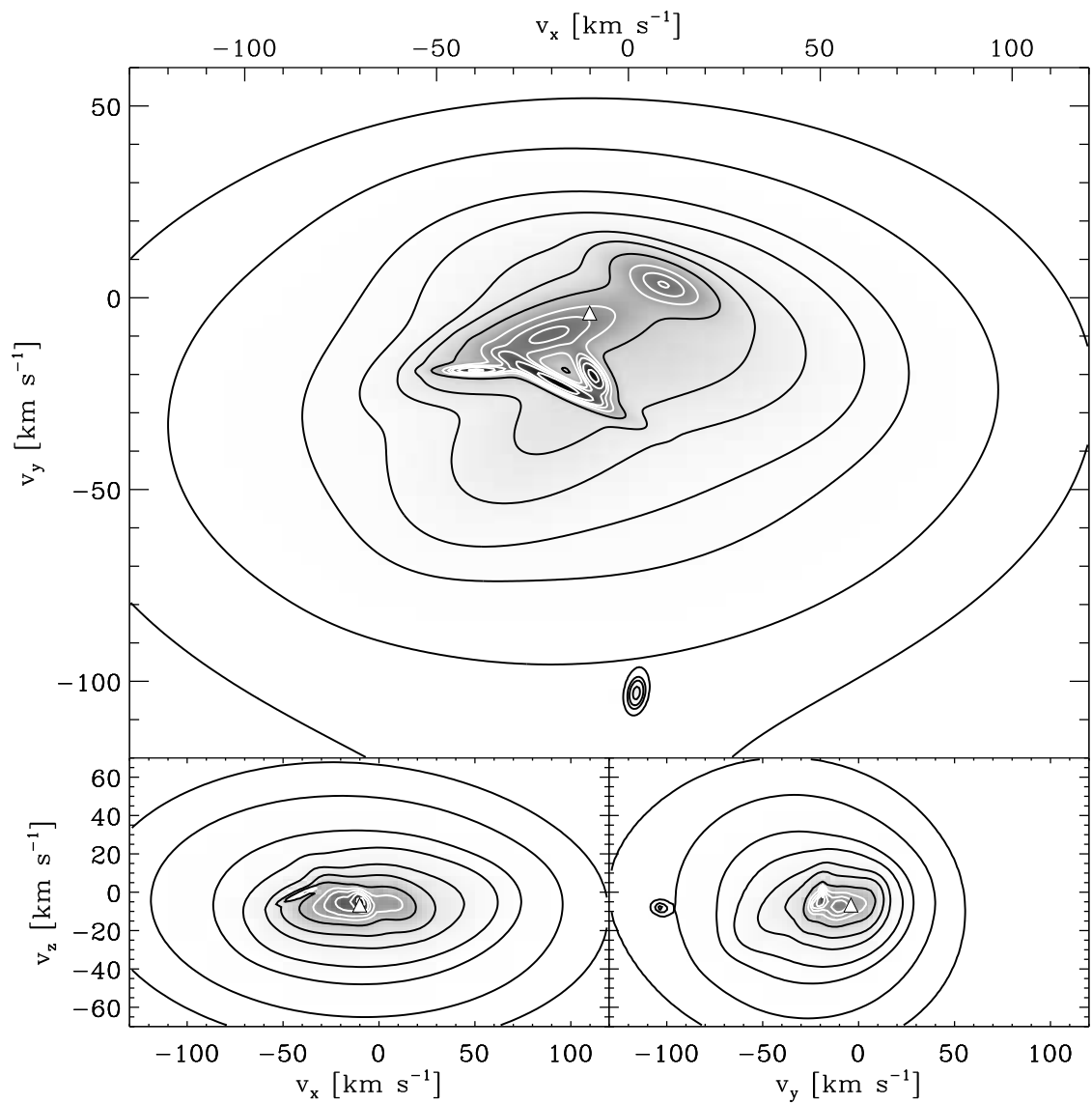}
\caption{Two-dimensional projections of the reconstructed velocity distribution with $K$ = 10 Gaussians and $w$ = 4 km$^2$ s$^{-2}$. Contours are as in \figurename~\ref{fig:veldensXY}. The origin is at the Solar velocity and the velocity of the Local Standard of Rest \citep{2005ApJ...629..268H} is indicated by a triangle.}%
\label{fig:annotated_veldist}
\end{figure}

We will discuss the properties of the reconstructed velocity
distribution in a separate paper (J.~Bovy, D.~W.~Hogg, \&
S.~T.~Roweis, 2009, in preparation), in which we will also look in
detail at the composition of the different clumps which we identify in
the velocity distribution, but here we briefly discuss the main
features of the velocity distribution shown in
\figurename~\ref{fig:annotated_veldist}.

Most of the features of the velocity distribution are in the
$\vx$--$\vy$ projection of the three-dimensional velocity
distribution. This is as expected, since phase mixing in the vertical
direction is much more efficient than in the horizontal direction
because the potential in the vertical direction varies more rapidly
with position than the potential in the horizontal direction in the
Solar neighborhood. The projection in the $\vx$--$\vz$ plane could be
well approximated by a single Gaussian distribution, as expected for a
well-mixed distribution of stars. The projection in the $\vy$--$\vz$
plane, except for a structure at large negative $\vy$ which we discuss
below, also corresponds to an approximately phase-mixed distribution
function, which is skewed because of the combined effect of the
exponential density profile of the stellar disk and the decreasing
velocity dispersion with Galactocentric radius in the disk
\citep{2008gady.book.....B}.

\begin{figure}
\includegraphics[height=0.717\textheight]{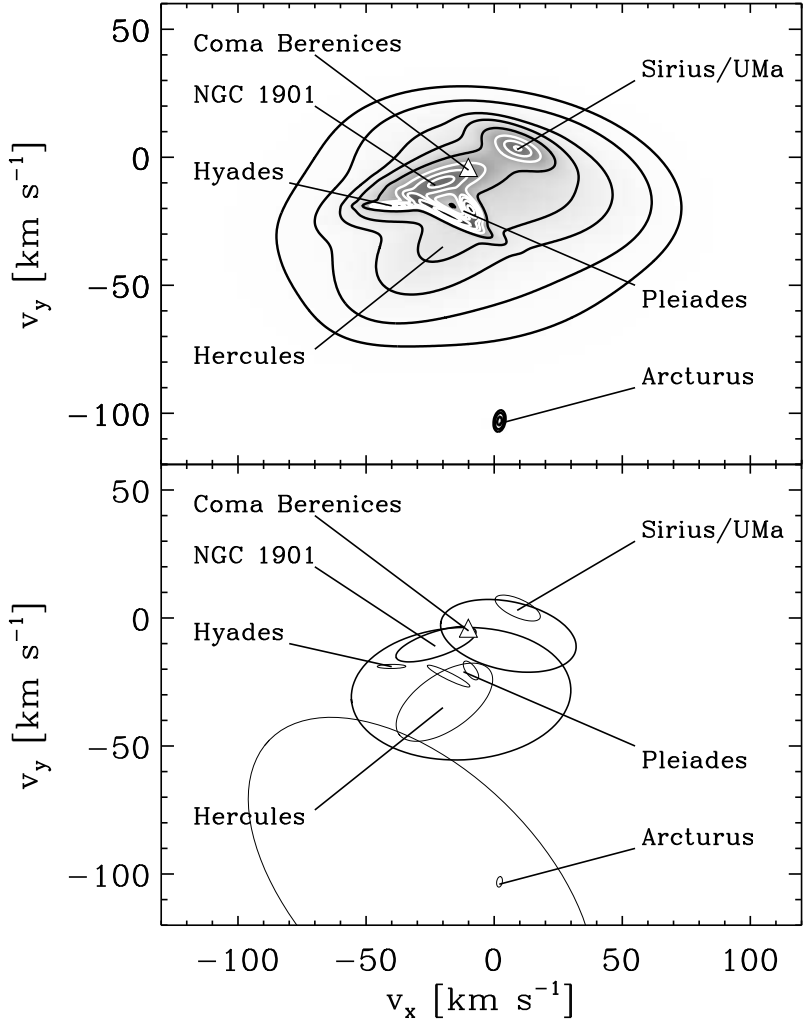}
\caption{Projection of the reconstructed velocity distribution with $K$ = 10 Gaussians and $w$ = 4 km$^2$ s$^{-2}$ in the $\vx$--$\vy$ plane: Velocity distribution with the moving groups indicated (\emph{top panel}); 1--sigma covariance ellipses around the mean of each Gaussian component $j$ with a linewidth proportional to the natural logarithm of its amplitude $\alpha_j$ (\emph{bottom panel}). Contours in the top panel are as in \figurename~\ref{fig:veldensXY}, but without the 99 and 99.9\,percent contours. The origin is at the Solar velocity and the velocity of the Local Standard of Rest \citep{2005ApJ...629..268H} is indicated by a triangle.}%
\label{fig:annotated_veldist2}
\end{figure}

In the $\vx$--$\vy$ plane we clearly see the most prominent moving
groups in their expected places. In
\figurename~\ref{fig:annotated_veldist2} we show the projection of the
velocity distribution in the $\vx$--$\vy$ plane with the moving groups
indicated. Also shown is a visual representation of the individual
Gaussian components of the velocity distribution in our
reconstruction: The location, covariance structure, and weight of each
component in the mixture is represented by 1--sigma contours around
the center of each Gaussian component with the linewidth indicating
the relative weight of the different components. Although we stress
that in the mixture of Gaussians approach to density esimation the
individual Gaussian components are meaningless, it is immediately
clear that there is a good and unambiguous correspondence between the
individual components and the modes of the distribution, which are
typically interpreted as moving groups. Therefore, in the following we
will use the Gaussian components as proxies for the peaks of the
distribution function. The Gaussian component with the largest weight
does not correspond to a peak in the velocity distribution and has a
large dispersion. Therefore, it is most probably part of the
background distribution. The moving group with the largest weight (as
judged by the amplitude of the corresponding Gaussian component in the
mixture, see Table~\ref{table:param}) is located at $\vx$ $\approx$
-23 km s$^{-1}$, $\vy$ $\approx$ -10 km s$^{-1}$, although it is
unclear at this point whether we can attribute all of the weight of
this component to the moving group or whether part of this component
should be identified with the background distribution of stars. This
group is known as NGC 1901 and it has been known for a long time. The
Coma Berenices moving group is typically found in this region as well;
we cannot naively associate a component of the mixture of Gaussians
with its fiducial location at $\vx$ $\approx$ -10 km s$^{-1}$, $\vy$
$\approx$ -5 km s$^{-1}$. However, looking at its location in
\figurename~\ref{fig:annotated_veldist2}, especially in the bottom
panel, we see that the region where the Coma Berenices moving group is
typically found is a part of the velocity distribution where three of
the Gaussian components with large weights in the mixture overlap. The
overdensity that this overlap gives rise to might indicate the
presence of the Coma Berenices moving group. We will come back to this
question in paper II, where we will perform a more sophisticated
analysis than naively associating components in the mixture with
moving groups of the velocity distribution and its substructures.

The Hercules group, which is not singled out by a contour in
\figurename~\ref{fig:annotated_veldist} but is nevertheless present in
our best fit model of the velocity distribution, is found at $\vx$
$\approx$ -20 km s$^{-1}$, $\vy$ $\approx$ -33 km s$^{-1}$. Whether
there is a valley between the Hercules group and the main extent of
the velocity distribution, is not entirely clear here, although it
does seem like the Hercules group merely sits on top of the background
distribution like all of the other moving groups. Following the
descending order of the weights of the Gaussian components, the Sirius
moving group, located at $\vx$ $\approx$ 9 km s$^{-1}$, $\vy$
$\approx$ 4 km s$^{-1}$ also stands out clearly at about the same
location that it has been reported at before
\citep[\eg][]{1998AJ....115.2384D}.

The Hyades and Pleiades moving groups are also clearly visible in the
velocity distribution: Two components of the mixture of Gaussians are
near where the Pleiades is traditionally located, at $\vx$ $\approx$
-15 km s$^{-1}$, $\vy$ $\approx$ -20 km s$^{-1}$. That there are two
components associated with the Pleiades moving group could mean that
there is substructure in the Pleiades moving group which has so far
gone unnoticed, or it could indicate that there is a dearth of stars
with velocities around $\vx$ $\approx$ -18 km s$^{-1}$, $\vy$
$\approx$ -19 km s$^{-1}$, forcing our fit of the velocity
distribution as a sum of Gaussian components to include two components
at the location of the Pleiades in order to reproduce the lack of
stars between the Pleiades moving group, the Hyades moving group, and
NGC 1901. More speculatively, this structure is reminiscent of the
kind of singularities that can appear when partially phase-mixed
structures are projected into a lower-dimensional space
\citep{1999MNRAS.307..877T}.

The Hyades moving group is also found at its expected location at
$\vx$ $\approx$ -40 km s$^{-1}$, $\vy$ $\approx$ -20 km
s$^{-1}$. Judging by the amplitude of the Gaussian component
corresponding to the Hyades moving group, this group only contains
about 1.75\,percent of the stars in the Solar neighborhood, much less
than the other moving groups discussed above.

We find one overdensity far removed from the main part of the velocity
distribution. This overdensity corresponds to the Arcturus stream, at
$\vx$ $\approx$ 0 km s$^{-1}$, $\vy$ $\approx$ -105 km s$^{-1}$. It
clearly stands out over the background of stars and can be seen in all
of our reconstructions of the velocity distribution with a sufficient
number of components, see \figurename~\ref{fig:veldensXY}. This group
of stars presumably belongs to the thick disk and it has been
hypothesized that it has an extragalactic origin
\citep{2004ApJ...601L..43N}. From
\figurename~\ref{fig:annotated_veldist} it is clear that it has a very
small width in velocity space, and judging by the covariance matrix of
the Gaussian component corresponding to the Arcturus group its
velocity dispersion is $\approx$ 2--3 km s$^{-1}$. This small velocity
dispersion casts doubts on the extragalactic interpretation of this
group. It is in sharp disagreement with the velocity dispersion of
$\approx$ 50 km s$^{-1}$ derived based on a sample of 46 stars
believed to be part of the Arcturus stream for which \Hipparcos\
parallaxes and proper motions are available
\citep{2004ApJ...601L..43N}. A similar small velocity dispersion as
found here has also been reported based on a joint analysis of the
\gcsabb\ data and a sample of metal-poor stars in the Solar
neighborhood \citep{2009IAUS..254..139W,2006A&A...445..939S}. The same
analysis also found that the Arcturus group is chemically similar to
the background thick disk stars, which is inconsistent with the
abundance properties of present-day dwarf Spheroidal galaxies, casting
further doubt on the interpretation of the Arcturus group as having an
extragalactic origin. We will discuss the properties of the Arcturus
group further in paper II.

Naively using the weights in the mixture of the components
corresponding to the different moving groups shows that many of these
groups contain up to 10\,percent of the stars in the Solar
neighborhood, although the large number of stars that seems to be part
of NGC 1901 is probably a consequence of contamination from background
stars for this moving group. The weights of the different moving
groups are in good agreement with the fractions of different moving
groups found by \citet{2005A&A...430..165F}, although we are able to
distinguish between more moving groups than was the case
earlier. Assuming that much of the second component of the mixture is
caused by the background leads us to the conclusion that $\approx$
40\,percent of the stars in the Solar neighborhood are part of a
moving group.

\section{Conclusion}

Our detailed analysis of the velocity distribution of nearby stars
from \Hipparcos\ data has lead us to the following conclusions:\\
$\bullet$ Performing a very conservative validation of our
reconstruction of the velocity distribution based on $\sim\!10,000$
tangential velocities of stars with the radial velocities of a sample
consisting of a similar number of stars shows that the amount of
complexity necessary in our model of the velocity distribution to
provide a good fit to the underlying velocity distribution is
relatively small. Adding more complexity to the model gives better
fits to the distribution of the tangential velocities, but fails to
provide a good fit to the validation set of radial velocities.
Therefore, unlike previous analyses, which validated their models of
the velocity distribution using only internal tests, we conclude that
there is not much evidence in the \Hipparcos\ data of much structure
in the velocity distribution beyond the major, known moving
groups. This does not preclude the existence of more structure in the
velocity distribution of nearby stars, but if more structure exists it
is present at only low significance in the \Hipparcos/\gcsabb\ data
set.\\ $\bullet$ Similarly, the same validation procedure shows that
the smallest scale on which we can reliably identify structure in the
velocity distribution from \Hipparcos\ data is a few km s$^{-1}$. This
calls into question claims from previous analyses of the \Hipparcos\
data of structure in the velocity distribution at the level of 1 km
s$^{-1}$. \\ $\bullet$ Our predictions of the radial velocities of
stars based on the model for the velocity distribution shows that we
get the bulk properties of the radial velocities correct using only
the tangential velocities of stars. This indicates that it is unlikely
that the addition of radial velocities to the sample on which we base
the fit would lead to a very different model for the velocity
distribution. This is unsurprising given the full-sky coverage of
\Hipparcos\ and the small spatial extent of our sample.\\$\bullet$
Predicting probability distributions for individual radial velocities
of stars based on the model of the velocity distribution and the
stars's tangential velocities and comparing them to the observed
radial velocities shows that there is still quite a bit of information
on average in the radial velocities: The predicted probability
distributions have relatively high entropy, \ie, they are not very
narrow in general, and the probability of the observed radial
velocities are rather small on average. In this context, each radial
velocity provides on average $\approx$ 6 bits of extra
information.\\$\bullet$ A preliminary investigation of the properties
of the velocity distribution shows that we recover all of the major
moving groups at the approximate locations at which they have been
found in the past: We find strong evidence for the Sirius/Ursa Major
group, the Hyades/Pleiades moving groups, the Hercules moving group,
the NGC 1901 group, and a thick disk stream, the Arcturus group. One
new feature that shows up in our reconstruction of the velocity
distribution is a dearth of stars between the Pleiades, the Hyades,
and the NGC 1901 moving groups, although it is unclear at the moment
whether the lack of stars in this region is a real underdensity or
whether it indicates substructure in the Pleiades moving group. A more
careful analysis of the reconstructed velocity distribution is
necessary to to distinguish between these two possibilities. We also
find that the Arcturus groups is kinematically cold, which calls its
interpretation as originating from the debris of a disrupted satellite
into question.\\ $\bullet$ A first look at the weights of the
different moving groups in the velocity distribution shows that about
40\,percent of the stars in the Solar neighborhood is part of a moving
group. Each of the major moving groups holds up to 10\,percent of the
stars.\\$\bullet$ As a result of our semi-parametric fit to the
velocity distribution, we have found a simple, explicit model for the
velocity distribution, given in Table~\ref{table:param}, which can be
used in theoretical studies and as the basis for a comparison of the
velocity distribution found here to that found by other methods.

\acknowledgments It is a pleasure to thank Fr{\'e}d{\'e}ric Arenou,
Michael Blanton, Anthony Brown, Walter Dehnen, Kathryn Johnston,
Dustin Lang, Floor van Leeuwen, Hans-Walter Rix, and Scott Tremaine
for comments and assistance. We also thank Walter Dehnen and James
Binney for use of their kinematically unbiased sample. JB and DWH were
partially supported by NASA (grant NNX08AJ48G). During part of the
period in which this research was performed, DWH was a research fellow
of the Alexander von Humboldt Foundation of Germany.

\clearpage
\begin{deluxetable}{cr@{.}lr@{.}lr@{.}lr@{.}lr@{.}lr@{.}lr@{.}lr@{.}lr@{.}lr@{.}l}
\tablecaption{Parameters of the component Gaussians for the density estimate with K = 10 and w = 4 km$^2$ s$^{-2}$.\label{table:param}}
\tablecolumns{21}
\tablewidth{0pt}
\rotate
\tablehead{\colhead{\ngauss} & \multicolumn{2}{c}{\alphagauss} & \multicolumn{2}{c}{$\eex\T\,\vv$} & \multicolumn{2}{c}{$\eey\T\,\vv$} & \multicolumn{2}{c}{$\eez\T\,\vv$} & \multicolumn{2}{c}{$\eex\T\,\VV\,\eex$} &\multicolumn{2}{c}{$\eey\T\,\VV\,\eey$} & \multicolumn{2}{c}{$\eez\T\,\VV\,\eez$} &\multicolumn{2}{c}{$\eex\T\,\VV\,\eey$} &\multicolumn{2}{c}{$\eex\T\,\VV\,\eez$} & \multicolumn{2}{c}{$\eey\T\,\VV\,\eez$} \\
\colhead{} & \multicolumn{2}{c}{} & \multicolumn{2}{c}{(km s$^{-1}$)} & \multicolumn{2}{c}{(km s$^{-1}$)} & \multicolumn{2}{c}{(km s$^{-1}$)} & \multicolumn{2}{c}{(km$^2$ s$^{-2}$)} & \multicolumn{2}{c}{(km$^2$ s$^{-2}$)} & \multicolumn{2}{c}{(km$^2$ s$^{-2}$)} & \multicolumn{2}{c}{(km$^2$ s$^{-2}$)} & \multicolumn{2}{c}{(km$^2$ s$^{-2}$)} & \multicolumn{2}{c}{(km$^2$ s$^{-2}$)}}
\startdata
           1 &            0 & 2392 &            5 & 54 &           -6 & 97 &           -9 & 26 &          699 & 85 &          202 & 09 &          143 & 63 &         -110 & 88 &           59 & 47 &           25 & 28\\
           2 &            0 & 2278 &          -22 & 72 &          -10 & 24 &           -7 & 30 &          243 & 91 &           47 & 60 &           40 & 45 &           72 & 78 &           10 & 75 &           10 & 35\\
           3 &            0 & 2262 &          -12 & 92 &          -29 & 57 &           -7 & 60 &         1836 & 31 &          670 & 79 &          537 & 94 &           56 & 82 &          -56 & 14 &          -31 & 09\\
           4 &            0 & 0891 &          -19 & 43 &          -32 & 90 &           -4 & 93 &          354 & 89 &          233 & 08 &          134 & 61 &          166 & 45 &          108 & 90 &          113 & 46\\
           5 &            0 & 0851 &            9 & 20 &            3 & 89 &           -5 & 99 &           77 & 82 &           25 & 69 &           48 & 58 &          -28 & 91 &          -18 & 81 &            8 & 87\\
           6 &            0 & 0674 &          -17 & 79 &          -22 & 69 &           -4 & 51 &           68 & 86 &           18 & 16 &           63 & 64 &          -32 & 89 &          -46 & 90 &           17 & 34\\
           7 &            0 & 0398 &           -9 & 07 &          -20 & 50 &           -4 & 88 &            9 & 04 &           13 & 47 &           14 & 04 &           -7 & 19 &           -5 & 56 &            7 & 63\\
           8 &            0 & 0174 &          -40 & 07 &          -18 & 92 &            0 & 64 &           30 & 99 &            0 & 64 &            6 & 14 &            0 & 25 &           13 & 15 &            0 & 19\\
           9 &            0 & 0072 &          -28 & 28 &         -105 & 62 &            2 & 87 &         4585 & 93 &         4465 & 62 &         3479 & 86 &        -2367 & 35 &         -328 & 67 &         -120 & 78\\
          10 &            0 & 0009 &            2 & 08 &         -103 & 07 &           -8 & 20 &            1 & 30 &            4 & 21 &            2 & 67 &            0 & 53 &            0 & 30 &            0 & 49\\
\enddata
\end{deluxetable}
\clearpage
\begin{deluxetable}{lrrrrrrrrrc}
\tablecaption{Low entropy predictions for Hipparcos stars' radial velocities\label{table:hip_low}}
\tablecolumns{11}
\tablewidth{0pt}
\tabletypesize{\small}
\tablehead{\colhead{ID\tablenotemark{\protect{\ref{HIPNUMBERtwo}}}} & \multicolumn{3}{c}{\ra\ (1991.25)}& \multicolumn{3}{c}{\dec\ (1991.25)}{} &\colhead{$v_r$\tablenotemark{\protect{\ref{VRtwo}}}}&\colhead{$v_r > (95\%)$\tablenotemark{\protect{\ref{VRLOWERtwo}}}}& \colhead{$v_r < (95\%)$\tablenotemark{\protect{\ref{VRUPPERtwo}}}} & \colhead{\phantom{bbb}$H$\tablenotemark{\protect{\ref{ENTROPYtwo}}}\phantom{bbb}}\\
\colhead{} & \colhead{(h} & \colhead{m} & \colhead{s)} & \colhead{($\degree$} & \colhead{$\prime$} & \colhead{$\prime\prime$)} & \colhead{(km s$^{-1}$)} & \colhead{(km s$^{-1}$)} & \colhead{(km s$^{-1}$)} & \colhead{}}
\startdata
HIP9044............&01&56&31.97&-60&13&37.6&4&-7&20&2.161\\
HIP66147..........&13&33&32.70&08&35&11.5&-13&-29&-10&2.352\\
HIP114271........&23&08&39.31&-15&03&09.4&-30&-37&-2&2.526\\
HIP110623........&22&24&37.33&16&53&48.8&-21&-25&0&2.577\\
HIP13976..........&03&00&02.62&07&44&58.9&32&24&43&2.679\\
HIP5542............&01&11&05.93&55&08&59.8&9&-22&21&2.696\\
HIP60529..........&12&24&30.09&31&16&49.1&3&-18&8&2.703\\
HIP171..............&00&02&09.65&27&05&04.2&-75&-81&-8&2.717\\
HIP57378..........&11&45&48.85&06&56&33.3&10&-11&12&2.731\\
HIP9065............&01&56&42.18&-49&24&27.5&51&1&55&2.735\\
HIP68801..........&14&05&03.83&10&00&48.5&-20&-38&-13&2.751\\
HIP115013........&23&17&37.80&-42&11&29.6&-14&-21&9&2.773\\
HIP73593..........&15&02&33.18&16&03&17.1&-27&-45&-10&2.776\\
HIP12158..........&02&36&41.57&-03&09&22.6&22&3&37&2.821\\
HIP77358..........&15&47&29.41&-37&54&56.9&-27&-45&2&2.834\\
HIP62724..........&12&51&12.17&19&09&40.8&-1&-24&2&2.839\\
HIP106856........&21&38&31.87&05&46&18.0&-29&-34&-4&2.852\\
HIP115151........&23&19&28.97&-70&19&23.1&-9&-23&26&2.864\\
HIP45320..........&09&14&10.67&-63&41&35.8&10&-6&48&2.869\\
HIP13511..........&02&54&00.97&-64&54&00.2&6&-9&34&2.889\\
HIP52332..........&10&41&40.91&-45&46&07.3&9&-16&38&2.891\\
HIP109182........&22&07&03.26&34&31&16.7&-19&-37&7&2.899\\
HIP70330..........&14&23&23.54&-27&49&19.3&46&-32&49&2.906\\
HIP6686............&01&25&48.60&60&14&07.5&10&-27&22&2.918\\
HIP88771..........&18&07&21.02&09&33&49.2&-38&-52&-1&2.959\\
\enddata
\tablecomments{Predictions for the radial velocity are based on the velocity density estimate using 10 Gaussians with regularization parameter $w$ = 4 km$^2$ s$^{-2}$, marginalized for the position on the sky of the star and conditioned on the stars' tangential velocity.}
\setcounter{tabletwo}{1}
\makeatletter
\let\@currentlabel\oldlabel
\newcommand{\@currentlabel}{\thetabletwo}
\makeatother
\renewcommand{\thetabletwo}{\alph{tabletwo}}
\tablenotetext{\thetabletwo}{\label{HIPNUMBERtwo}
\Hipparcos\ number.\stepcounter{tabletwo}}
\tablenotetext{\thetabletwo}{\label{VRtwo}
Maximum likelihood estimate of the radial velocity.\stepcounter{tabletwo}}
\tablenotetext{\thetabletwo}{\label{VRLOWERtwo}
Lower limit on the radial velocity (95-percent confidence).\stepcounter{tabletwo}}
\tablenotetext{\thetabletwo}{\label{VRUPPERtwo}
Upper limit on the radial velocity (95-percent confidence).\stepcounter{tabletwo}}
\tablenotetext{\thetabletwo}{\label{ENTROPYtwo}
Entropy of the probability distribution of the radial velocity.\stepcounter{tabletwo}}
\end{deluxetable}
\clearpage
\begin{deluxetable}{lrrrrrrrrrc}
\tablecaption{High entropy predictions for Hipparcos stars' radial velocities\label{table:hip_high}}
\tablecolumns{11}
\tablewidth{0pt}
\tabletypesize{\small}
\tablehead{\colhead{ID\tablenotemark{\protect{\ref{HIPNUMBERone}}}} & \multicolumn{3}{c}{\ra\ (1991.25)}& \multicolumn{3}{c}{\dec\ (1991.25)}{} &\colhead{$v_r$\tablenotemark{\protect{\ref{VRone}}}}&\colhead{$v_r > (95\%)$\tablenotemark{\protect{\ref{VRLOWERone}}}}& \colhead{$v_r < (95\%)$\tablenotemark{\protect{\ref{VRUPPERone}}}} & \colhead{\phantom{bbb}$H$\tablenotemark{\protect{\ref{ENTROPYone}}}\phantom{bbb}}\\
\colhead{} & \colhead{(h} & \colhead{m} & \colhead{s)} & \colhead{($\degree$} & \colhead{$\prime$} & \colhead{$\prime\prime$)} & \colhead{(km s$^{-1}$)} & \colhead{(km s$^{-1}$)} & \colhead{(km s$^{-1}$)} & \colhead{}}
\startdata
HIP67534..........&13&50&16.63&-57&15&40.3&57&-92&209&5.817\\
HIP64444..........&13&12&32.07&-34&44&50.6&72&-65&210&5.717\\
HIP95344..........&19&23&49.32&-81&32&24.9&47&-61&216&5.717\\
HIP117029........&23&43&26.61&58&04&44.9&-46&-205&71&5.694\\
HIP84164..........&17&12&21.95&-46&33&40.0&40&-100&173&5.663\\
HIP110618........&22&24&34.39&-72&15&13.6&-37&-142&126&5.639\\
HIP117702........&23&52&14.21&-61&25&23.4&61&-68&184&5.589\\
HIP7869............&01&41&14.20&-67&40&33.1&80&-44&198&5.555\\
HIP117254........&23&46&32.69&-41&34&47.3&25&-98&142&5.529\\
HIP63559..........&13&01&26.58&-27&22&26.5&-4&-74&171&5.503\\
HIP102862........&20&50&22.62&-40&36&26.6&-9&-136&99&5.503\\
HIP55988..........&11&28&27.91&07&31&12.9&61&-57&173&5.493\\
HIP98532..........&20&01&00.43&-12&15&17.1&-73&-176&62&5.493\\
HIP111286........&22&32&40.70&-65&26&07.3&39&-73&171&5.465\\
HIP76976..........&15&43&03.76&-10&55&57.9&44&-70&152&5.451\\
HIP5549............&01&11&10.39&-82&32&54.0&27&-47&185&5.432\\
HIP3086............&00&39&12.81&03&07&59.5&13&-130&97&5.423\\
HIP78400..........&16&00&19.98&-16&31&56.6&-29&-119&92&5.395\\
HIP17671..........&03&47&06.05&-56&02&29.5&42&-58&152&5.384\\
HIP107731........&21&49&24.76&-47&58&35.9&19&-112&103&5.337\\
HIP96185..........&19&33&27.40&33&12&04.8&-49&-185&7&5.336\\
HIP99224..........&20&08&33.81&-15&43&43.7&-38&-148&43&5.302\\
HIP100792........&20&26&11.85&09&27&05.2&-88&-185&4&5.293\\
HIP30439..........&06&23&57.61&-45&56&52.2&60&-11&175&5.284\\
HIP86013..........&17&34&43.34&06&00&48.3&-87&-182&1&5.261\\
\enddata
\tablecomments{Predictions for the radial velocity are based on the velocity density estimate using 10 Gaussians with regularization parameter $w$ = 4 km$^2$ s$^{-2}$, marginalized for the position on the sky of the star and conditioned on the stars' tangential velocity.}
\setcounter{tableone}{1}
\makeatletter
\let\@currentlabel\oldlabel
\newcommand{\@currentlabel}{\thetableone}
\makeatother
\renewcommand{\thetableone}{\alph{tableone}}
\tablenotetext{\thetableone}{\label{HIPNUMBERone}
\Hipparcos\ number.\stepcounter{tableone}}
\tablenotetext{\thetableone}{\label{VRone}
Maximum likelihood estimate of the radial velocity.\stepcounter{tableone}}
\tablenotetext{\thetableone}{\label{VRLOWERone}
Lower limit on the radial velocity (95-percent confidence).\stepcounter{tableone}}
\tablenotetext{\thetableone}{\label{VRUPPERone}
Upper limit on the radial velocity (95-percent confidence).\stepcounter{tableone}}
\tablenotetext{\thetableone}{\label{ENTROPYone}
Entropy of the probability distribution of the radial velocity.\stepcounter{tableone}}
\end{deluxetable}

\clearpage

\clearpage

\clearpage

\clearpage

\clearpage

\clearpage
\includegraphics{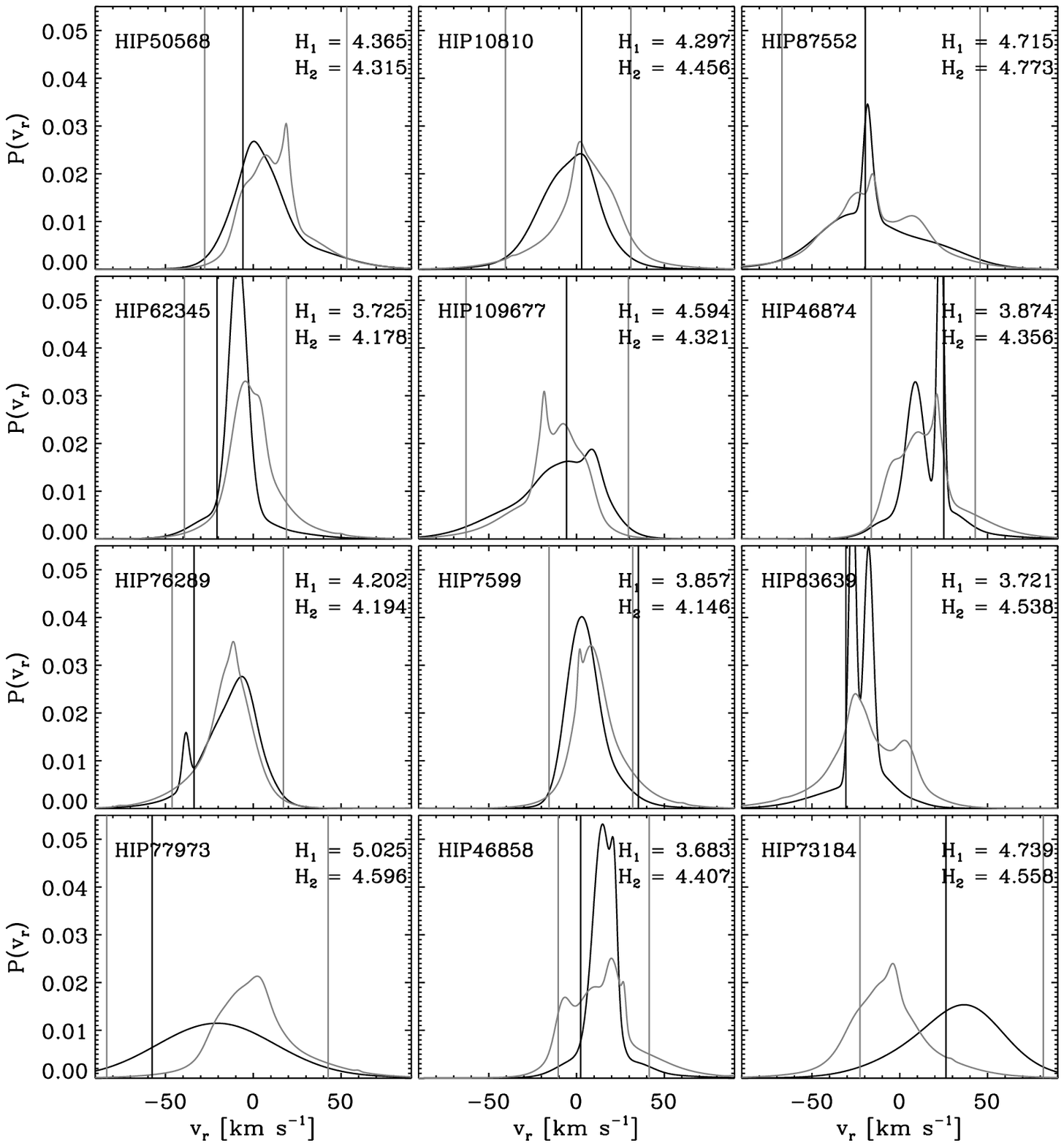}
\clearpage
\begin{figure}
\caption{Predicted radial velocity distribution using the reconstructed velocity distribution with $k$ = 10, $w$ = 4 km$^2$ s$^{-1}$ for random stars in the \gcsabb\ catalogue. The gray curve gives the radial velocity distribution obtained by marginalizing the three-dimensional velocity distribution over the tangential velocity (see \eqnnumber~[\ref{eq:margpredicted}]), while the black curve shows the radial velocity distribution obtained by conditioning the reconstructed velocity distribution on the tangential velocity of the star as measured by \Hipparcos (see \eqnnumber~[\ref{eq:condpredicted}]). The black vertical line gives the measured value of the radial velocity from the \gcsabb\ catalogue. The gray vertical lines give the 95-percent confidence interval limits for the conditional distribution. The entropies $H_1$ and $H_2$ of the conditional, respectively marginalized distribution are given in the upper-right corner. The \Hipparcos\ number of the star is given in the upper-left corner.}%
\label{fig:predict_gcs_rand}
\end{figure}

\clearpage
\begin{figure}
\includegraphics{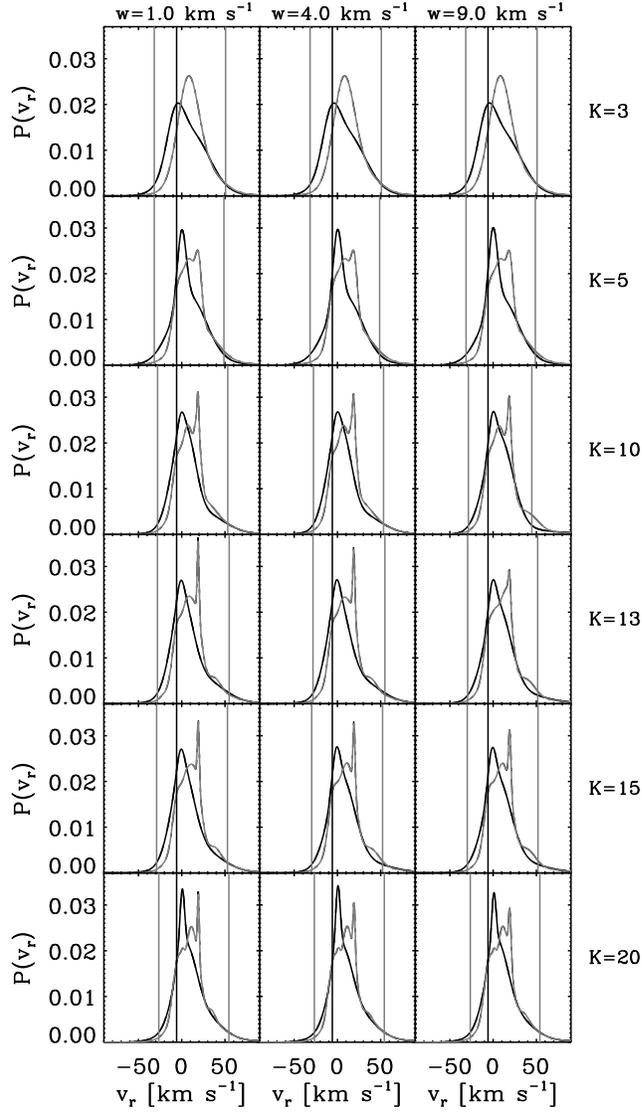}
\caption{Predicted radial velocity distribution using the reconstructed velocity distribution as a function of the number of Gaussians $K$ used in the reconstruction and the regularization parameter $w$ for a random star in the \gcsabb\ catalogue (the star with \Hipparcos\ number HIP50568). See Fig.~\ref{fig:predict_gcs_rand} for an explanation of the different lines in each panel.}%
\label{fig:tile_onepred}
\end{figure}

\clearpage

\clearpage

\clearpage

\clearpage

\clearpage

\clearpage

\clearpage

\clearpage

\clearpage

\clearpage

\clearpage

\clearpage

\clearpage

\clearpage


\begin{thebibliography}{118}
\expandafter\ifx\csname natexlab\endcsname\relax\def\natexlab#1{#1}\fi

\bibitem[{{Akaike}(1974)}]{Akaike}
{Akaike}, H. 1974, {IEEE Transactions on Automatic Control}, 19, 716

\bibitem[{{Antoja} {et~al.}(2008){Antoja}, {Figueras}, {Fern{\'a}ndez}, \&
  {Torra}}]{2008A&A...490..135A}
{Antoja}, T., {Figueras}, F., {Fern{\'a}ndez}, D., \& {Torra}, J. 2008, \aap,
  490, 135

\bibitem[{{Asiain} {et~al.}(1999){Asiain}, {Figueras}, {Torra}, \&
  {Chen}}]{1999A&A...341..427A}
{Asiain}, R., {Figueras}, F., {Torra}, J., \& {Chen}, B. 1999, \aap, 341, 427

\bibitem[{{Baade}(1944)}]{1944ApJ...100..137B}
{Baade}, W. 1944, \apj, 100, 137

\bibitem[{{Baade}(1958)}]{baade1958a}
{Baade}, W. 1958, in {Stellar Populations: Proceedings of the conference
  sponsored by the Pontifical Academy of Science and the Vatican Observatory,
  May 20-28, 1957}, ed. D.~J.~K. {O'Connell} ({Amsterdam}: {North-Holland
  Pub.~Co.}), 3

\bibitem[{{Bahcall} \& {Soneira}(1980)}]{1980ApJS...44...73B}
{Bahcall}, J.~N. \& {Soneira}, R.~M. 1980, \apjs, 44, 73

\bibitem[{{Bahcall} \& {Soneira}(1984)}]{1984ApJS...55...67B}
{Bahcall}, J.~N. \& {Soneira}, R.~M. 1984, \apjs, 55, 67

\bibitem[{{Bienaym{\'e}}(1999)}]{1999A&A...341...86B}
{Bienaym{\'e}}, O. 1999, \aap, 341, 86

\bibitem[{{Binney} \& {Tremaine}(2008)}]{2008gady.book.....B}
{Binney}, J. \& {Tremaine}, S. 2008, {Galactic Dynamics: Second Edition}
  (Princeton University Press)

\bibitem[{{Binney} {et~al.}(1997){Binney}, {Dehnen}, {Houk}, {Murray}, \&
  {Penston}}]{1997ESASP.402..473B}
{Binney}, J.~J., {Dehnen}, W., {Houk}, N., {Murray}, C.~A., \& {Penston}, M.~J.
  1997, in ESA Special Publication, Vol. 402, Hipparcos - Venice '97, 473

\bibitem[{{Blaauw} {et~al.}(1960){Blaauw}, {Gum}, {Pawsey}, \&
  {Westerhout}}]{1960MNRAS.121..123B}
{Blaauw}, A., {Gum}, C.~S., {Pawsey}, J.~L., \& {Westerhout}, G. 1960, \mnras,
  121, 123

\bibitem[{{Boesgaard} \& {Budge}(1988)}]{1988ApJ...332..410B}
{Boesgaard}, A.~M. \& {Budge}, K.~G. 1988, \apj, 332, 410

\bibitem[{{Boss}(1911)}]{1911AJ.....27...33B}
{Boss}, B. 1911, \aj, 27, 33

\bibitem[{{Boss}(1908)}]{boss08a}
{Boss}, L. 1908, \aj, 26, 31

\bibitem[{{Bovy} {et~al.}(2009){Bovy}, {Hogg}, \& {Roweis}}]{Bovy09a}
{Bovy}, J., {Hogg}, D.~W., \& {Roweis}, S.~T. 2009, {arXiv:0905.2979 [stat.ME]}

\bibitem[{{Boyle} \& {McClure}(1975)}]{1975PASP...87...17B}
{Boyle}, R.~J. \& {McClure}, R.~D. 1975, \pasp, 87, 17

\bibitem[{{Bozdogan}(1983)}]{Bozdogan}
{Bozdogan}, H. 1983, {Determining the number of component clusters in the
  standard multivariate normal mixture model using model-selection criteria},
  Tech. rep., {TR UIC/DQM/A83-1, Quantitative Methods Dept., University of
  Illinois}

\bibitem[{{Breger}(1968)}]{1968PASP...80..578B}
{Breger}, M. 1968, \pasp, 80, 578

\bibitem[{{Cabrera-Ca\~{n}o} \& {Alfaro}(1990)}]{1990A&A...235...94C}
{Cabrera-Ca\~{n}o}, J. \& {Alfaro}, E.~J. 1990, \aap, 235, 94

\bibitem[{{Chaudhuri}(1940)}]{1940MNRAS.100..574C}
{Chaudhuri}, P.~C. 1940, \mnras, 100, 574

\bibitem[{{Chen} {et~al.}(1997){Chen}, {Asiain}, {Figueras}, \&
  {Torra}}]{1997A&A...318...29C}
{Chen}, B., {Asiain}, R., {Figueras}, F., \& {Torra}, J. 1997, \aap, 318, 29

\bibitem[{{Chereul} {et~al.}(1998){Chereul}, {Creze}, \&
  {Bienayme}}]{1998A&A...340..384C}
{Chereul}, E., {Creze}, M., \& {Bienayme}, O. 1998, \aap, 340, 384

\bibitem[{{Chereul} {et~al.}(1999){Chereul}, {Cr{\'e}z{\'e}}, \&
  {Bienaym{\'e}}}]{1999A&AS..135....5C}
{Chereul}, E., {Cr{\'e}z{\'e}}, M., \& {Bienaym{\'e}}, O. 1999, \aaps, 135, 5

\bibitem[{{Conway} \& {Sloane}(1992)}]{conway92a}
{Conway}, J.~H. \& {Sloane}, N. J.~A. 1992, {Sphere Packings, Lattices and
  Groups} (London: {Springer-Verlag})

\bibitem[{{Dehnen}(1998)}]{1998AJ....115.2384D}
{Dehnen}, W. 1998, \aj, 115, 2384

\bibitem[{{Dehnen} \& {Binney}(1998)}]{1998MNRAS.298..387D}
{Dehnen}, W. \& {Binney}, J.~J. 1998, \mnras, 298, 387

\bibitem[{{Dempster} {et~al.}(1977){Dempster}, {Laird}, \&
  {Rubin}}]{Dempster1977}
{Dempster}, A.~P., {Laird}, N.~M., \& {Rubin}, D.~B. 1977, Journal of the Royal
  Statistical Society. Series B (Methodological), 39, 1

\bibitem[{{Eddington}(1910)}]{1910MNRAS..71...43E}
{Eddington}, A.~S. 1910, \mnras, 71, 43

\bibitem[{{Eggen}(1958)}]{1958MNRAS.118..154E}
{Eggen}, O.~J. 1958, \mnras, 118, 154

\bibitem[{{Eggen}(1959{\natexlab{a}})}]{1959Obs....79..182E}
{Eggen}, O.~J. 1959{\natexlab{a}}, The Observatory, 79, 182

\bibitem[{{Eggen}(1959{\natexlab{b}})}]{1959Obs....79...88E}
{Eggen}, O.~J. 1959{\natexlab{b}}, The Observatory, 79, 88

\bibitem[{{Eggen}(1964)}]{1964RGOB...84..111E}
{Eggen}, O.~J. 1964, Roy.~Greenwich Obs.~Bull., 84, 111

\bibitem[{{Eggen}(1965)}]{1965Obs....85..191E}
{Eggen}, O.~J. 1965, The Observatory, 85, 191

\bibitem[{{Eggen}(1969)}]{1969PASP...81..553E}
{Eggen}, O.~J. 1969, \pasp, 81, 553

\bibitem[{{Eggen}(1970)}]{1970PASP...82...99E}
{Eggen}, O.~J. 1970, \pasp, 82, 99

\bibitem[{{Eggen}(1971{\natexlab{a}})}]{1971PASP...83..271E}
{Eggen}, O.~J. 1971{\natexlab{a}}, \pasp, 83, 271

\bibitem[{{Eggen}(1971{\natexlab{b}})}]{1971PASP...83..251E}
{Eggen}, O.~J. 1971{\natexlab{b}}, \pasp, 83, 251

\bibitem[{{Eggen}(1978)}]{1978ApJ...222..203E}
{Eggen}, O.~J. 1978, \apj, 222, 203

\bibitem[{{Eggen}(1983)}]{1983AJ.....88..813E}
{Eggen}, O.~J. 1983, \aj, 88, 813

\bibitem[{{Eggen}(1986)}]{1986AJ.....92..910E}
{Eggen}, O.~J. 1986, \aj, 92, 910

\bibitem[{{Eggen} \& {Sandage}(1959)}]{1959MNRAS.119..255E}
{Eggen}, O.~J. \& {Sandage}, A.~R. 1959, \mnras, 119, 255

\bibitem[{{ESA}(1997)}]{ESA97a}
{ESA}. 1997, {The \emph{Hipparcos} and Tycho Catalogues} (Noordwijk: ESA: {ESA
  SP-1200})

\bibitem[{{Famaey} {et~al.}(2005){Famaey}, {Jorissen}, {Luri}, {Mayor}, {Udry},
  {Dejonghe}, \& {Turon}}]{2005A&A...430..165F}
{Famaey}, B., {Jorissen}, A., {Luri}, X., {Mayor}, M., {Udry}, S., {Dejonghe},
  H., \& {Turon}, C. 2005, \aap, 430, 165

\bibitem[{{Figueras} {et~al.}(1997){Figueras}, {Gomez}, {Asiain}, {Chen},
  {Comeron}, {Grenier}, {Lebreton}, {Moreno}, {Sabas}, \&
  {Torra}}]{1997ESASP.402..519F}
{Figueras}, F., {Gomez}, A.~E., {Asiain}, R., {Chen}, B., {Comeron}, F.,
  {Grenier}, S., {Lebreton}, Y., {Moreno}, M., {Sabas}, V., \& {Torra}, J.
  1997, in ESA Special Publication, Vol. 402, Hipparcos - Venice '97, 519--524

\bibitem[{{Francis} \& {Anderson}(2009)}]{2009arXiv0901.3503F}
{Francis}, C. \& {Anderson}, E. 2009, arXiv:0901.3503 [astro-ph]

\bibitem[{{Gelman} {et~al.}(2000){Gelman}, {Carlin}, {Stern}, \&
  {Rubin}}]{Gelman00a}
{Gelman}, A., {Carlin}, J.~B., {Stern}, H.~S., \& {Rubin}, D.~B. 2000,
  {Bayesian Data Analysis} ({Chapman \& Hall/CRC})

\bibitem[{{Gerbaldi} {et~al.}(1989)}]{1989Msngr..56...12G}
{Gerbaldi}, M. {et~al.} 1989, ESO Messenger, 56, 12

\bibitem[{{Ghigna} {et~al.}(1998){Ghigna}, {Moore}, {Governato}, {Lake},
  {Quinn}, \& {Stadel}}]{ghigna98a}
{Ghigna}, S., {Moore}, B., {Governato}, F., {Lake}, G., {Quinn}, T., \&
  {Stadel}, J. 1998, \mnras, 300, 146

\bibitem[{{Gilmore} \& {Reid}(1983)}]{1983MNRAS.202.1025G}
{Gilmore}, G. \& {Reid}, N. 1983, \mnras, 202, 1025

\bibitem[{{Gomez} {et~al.}(1990){Gomez}, {Delhaye}, {Grenier}, {Jaschek},
  {Arenou}, \& {Jaschek}}]{1990A&A...236...95G}
{Gomez}, A.~E., {Delhaye}, J., {Grenier}, S., {Jaschek}, C., {Arenou}, F., \&
  {Jaschek}, M. 1990, \aap, 236, 95

\bibitem[{{Grenier} {et~al.}(1985){Grenier}, {Gomez}, {Jaschek}, {Jaschek}, \&
  {Heck}}]{1985A&A...145..331G}
{Grenier}, S., {Gomez}, A.~E., {Jaschek}, C., {Jaschek}, M., \& {Heck}, A.
  1985, \aap, 145, 331

\bibitem[{{Gr{\"u}nwald}(2007)}]{Grunwaldbook}
{Gr{\"u}nwald}, P.~D. 2007, {The minimum description length principle}
  (Cambridge, Massachusetts: {MIT Press})

\bibitem[{{Helmi} {et~al.}(2003){Helmi}, {White}, \&
  {Springel}}]{2003MNRAS.339..834H}
{Helmi}, A., {White}, S.~D.~M., \& {Springel}, V. 2003, \mnras, 339, 834

\bibitem[{{Hertzsprung}(1909)}]{1909ApJ....30..135H}
{Hertzsprung}, E. 1909, \apj, 30, 135

\bibitem[{{Hogg} {et~al.}(2005){Hogg}, {Blanton}, {Roweis}, \&
  {Johnston}}]{2005ApJ...629..268H}
{Hogg}, D.~W., {Blanton}, M.~R., {Roweis}, S.~T., \& {Johnston}, K.~V. 2005,
  \apj, 629, 268

\bibitem[{{Ivezi{\'c}} {et~al.}(2008)}]{2008ApJ...684..287I}
{Ivezi{\'c}}, {\v Z}. {et~al.} 2008, \apj, 684, 287

\bibitem[{{Johnston}(1998)}]{johnston98a}
{Johnston}, K.~V. 1998, \apj, 495, 297

\bibitem[{{Kapteyn}(1905)}]{kapteyn05a}
{Kapteyn}, J.~C. 1905, {British Assoc.~Adv.~Sci.~Rep.}, Sec.~A, 257

\bibitem[{{Kapteyn}(1914)}]{1914ApJ....40...43K}
{Kapteyn}, J.~C. 1914, \apj, 40, 43

\bibitem[{{Kolmogorov}(1965)}]{kolmogorov65a}
{Kolmogorov}, A.~N. 1965, {Problems.~Inform.~Transmission}, 1, 1

\bibitem[{{Koposov} {et~al.}(2009){Koposov}, {Yoo}, {Rix}, {Weinberg},
  {Macci{\`o}}, \& {Escud{\'e}}}]{Koposov:2009ru}
{Koposov}, S.~E., {Yoo}, J., {Rix}, H.-W., {Weinberg}, D.~H., {Macci{\`o}},
  A.~V., \& {Escud{\'e}}, J.~M. 2009, \apj, 696, 2179

\bibitem[{{Koposov} {et~al.}(2008)}]{koposov}
{Koposov}, S.~E. {et~al.} 2008, \apj, 686, 279

\bibitem[{{Luri} {et~al.}(1996){Luri}, {Mennessier}, {Torra}, \&
  {Figueras}}]{1996A&AS..117..405L}
{Luri}, X., {Mennessier}, M.~O., {Torra}, J., \& {Figueras}, F. 1996, \aaps,
  117, 405

\bibitem[{{M{\"a}dler}(1846)}]{1846AN.....24..213M}
{M{\"a}dler}, J.~H. 1846, AN, 24, 213

\bibitem[{{M{\"a}dler}(1847)}]{madler47}
{M{\"a}dler}, J.~H. 1847, {Untersuchungen {\"u}ber die Fixstern-systeme}
  (Lepizig: Mitau)

\bibitem[{{McDonald} \& {Hearnshaw}(1983)}]{1983MNRAS.204..841M}
{McDonald}, A.~R.~E. \& {Hearnshaw}, J.~B. 1983, \mnras, 204, 841

\bibitem[{{Murenzi}(1989)}]{1989wtfm.conf..239M}
{Murenzi}, R. 1989, in Wavelets. Time-Frequency Methods and Phase Space, ed.
  J.-M. {Combes}, A.~{Grossmann}, \& P.~{Tchamitchian}, 239--+

\bibitem[{{Navarro} {et~al.}(2004){Navarro}, {Helmi}, \&
  {Freeman}}]{2004ApJ...601L..43N}
{Navarro}, J.~F., {Helmi}, A., \& {Freeman}, K.~C. 2004, \apjl, 601, L43

\bibitem[{{Nordstr{\"o}m} {et~al.}(2004){Nordstr{\"o}m}, {Mayor}, {Andersen},
  {Holmberg}, {Pont}, {J{\o}rgensen}, {Olsen}, {Udry}, \&
  {Mowlavi}}]{2004A&A...418..989N}
{Nordstr{\"o}m}, B., {Mayor}, M., {Andersen}, J., {Holmberg}, J., {Pont}, F.,
  {J{\o}rgensen}, B.~R., {Olsen}, E.~H., {Udry}, S., \& {Mowlavi}, N. 2004,
  \aap, 418, 989

\bibitem[{{Ogorodnikov} \& {Latyshev}(1968)}]{1968SvA....12..279O}
{Ogorodnikov}, K.~F. \& {Latyshev}, I.~N. 1968, Soviet Astronomy, 12, 279

\bibitem[{{Ogorodnikov} \& {Latyshev}(1970)}]{1970SvA....13..934O}
{Ogorodnikov}, K.~F. \& {Latyshev}, I.~N. 1970, Soviet Astronomy, 13, 934

\bibitem[{{Oliver} \& {Baxter}(1994)}]{oliver94a}
{Oliver}, J.~J. \& {Baxter}, R.~A. 1994, {MML and Bayesianism: Similarities and
  Differences}, Tech. Rep. Tech Report 206, Dept. of Computer Science, Monash
  University, Clayton, Vic. 3168, Australia

\bibitem[{{Oliver} {et~al.}(1996){Oliver}, {Baxter}, \& {Wallace}}]{Oliver96a}
{Oliver}, J.~J.~O., {Baxter}, R.~A., \& {Wallace}, C.~S. 1996, in In Machine
  Learning: Proceedings of the Thirteenth International Conference (ICML 96
  (Morgan Kaufmann Publishers), 364

\bibitem[{{Oort}(1958)}]{oort58a}
{Oort}, J.~H. 1958, in {Stellar Populations: Proceedings of the conference
  sponsored by the Pontifical Academy of Science and the Vatican Observatory,
  May 20-28, 1957}, ed. D.~J.~K. {O'Connell} ({Amsterdam}: {North-Holland
  Pub.~Co.}), 515

\bibitem[{{Ormoneit} \& {Tresp}(1996)}]{Ormoneit1995}
{Ormoneit}, D. \& {Tresp}, V. 1996, in {Advances in Neural Information
  Processing Systems 8, NIPS, Denver,CO, November 27-30, 1995}, ed. D.~S.
  {Touretzky}, M.~{Mozer}, \& M.~E. {Hasselmo} (MIT Press)

\bibitem[{{Perryman} {et~al.}(2001)}]{2001A&A...369..339P}
{Perryman}, M.~A.~C. {et~al.} 2001, \aap, 369, 339

\bibitem[{{Plummer}(1913)}]{1913MNRAS..73..492P}
{Plummer}, H.~C. 1913, \mnras, 73, 492

\bibitem[{{Proctor}(1869)}]{proctor69a}
{Proctor}, R.~A. 1869, Proc.~Roy.~Soc.~London, 18, 169

\bibitem[{{Proust} \& {Foy}(1988)}]{1988Ap&SS.145...61P}
{Proust}, D. \& {Foy}, R. 1988, \apss, 145, 61

\bibitem[{{Rasmuson}(1921)}]{rasmuson21}
{Rasmuson}, N.~H. 1921, Med.~Lunds.~Obs., {Ser.~II}, {No.~26}

\bibitem[{{Rissanen}(1978)}]{rissanen78a}
{Rissanen}, J. 1978, {Automatica}, 14, 465

\bibitem[{{Roberts} {et~al.}(1998){Roberts}, {Husmeier}, {Rezek}, \&
  {Penny}}]{roberts98a}
{Roberts}, S.~J., {Husmeier}, D., {Rezek}, I., \& {Penny}, W. 1998, IEEE
  Transactions on Pattern Analysis and Machine Intelligence, 20, 1133

\bibitem[{{Roman}(1949)}]{1949ApJ...110..205R}
{Roman}, N.~G. 1949, \apj, 110, 205

\bibitem[{{Roman}(1954)}]{1954AJ.....59..307R}
{Roman}, N.~G. 1954, \aj, 59, 307

\bibitem[{{Russell}(1912)}]{1912AJ.....27...96R}
{Russell}, H.~N. 1912, \aj, 27, 96

\bibitem[{{Schuster} {et~al.}(2006){Schuster}, {Moitinho}, {M{\'a}rquez},
  {Parrao}, \& {Covarrubias}}]{2006A&A...445..939S}
{Schuster}, W.~J., {Moitinho}, A., {M{\'a}rquez}, A., {Parrao}, L., \&
  {Covarrubias}, E. 2006, \aap, 445, 939

\bibitem[{{Schwarz}(1978)}]{schwarz78a}
{Schwarz}, G. 1978, {Ann.~Stat.}, 6, 461

\bibitem[{{Schwarzschild}(1907)}]{schwarzschild07a}
{Schwarzschild}, K. 1907, {Nachrichten von der K{\"o}niglichen Gesellschaft der
  Wissenschaften zu G{\"o}ttingen}, 5, {614}

\bibitem[{{Schwarzschild}(1958)}]{schwarzschild58a}
{Schwarzschild}, M. 1958, in {Stellar Populations: Proceedings of the
  conference sponsored by the Pontifical Academy of Science and the Vatican
  Observatory, May 20-28, 1957}, ed. D.~J.~K. {O'Connell} ({Amsterdam}:
  {North-Holland Pub.~Co.}), 207

\bibitem[{{Silverman}(1986)}]{Silverman86a}
{Silverman}, B.~W. 1986, {Density Estimation for Statistics and Data Analysis}
  ({Chapman and Hall})

\bibitem[{{Simon} \& {Geha}(2007)}]{2007ApJ...670..313S}
{Simon}, J.~D. \& {Geha}, M. 2007, \apj, 670, 313

\bibitem[{{Skuljan} {et~al.}(1999){Skuljan}, {Hearnshaw}, \&
  {Cottrell}}]{1999MNRAS.308..731S}
{Skuljan}, J., {Hearnshaw}, J.~B., \& {Cottrell}, P.~L. 1999, \mnras, 308, 731

\bibitem[{{Slezak} {et~al.}(1990){Slezak}, {Bijaoui}, \&
  {Mars}}]{1990A&A...227..301S}
{Slezak}, E., {Bijaoui}, A., \& {Mars}, G. 1990, \aap, 227, 301

\bibitem[{{Soderblom} \& {Clements}(1987)}]{1987AJ.....93..920S}
{Soderblom}, D.~R. \& {Clements}, S.~D. 1987, \aj, 93, 920

\bibitem[{{Soderblom} \& {Mayor}(1993)}]{1993AJ....105..226S}
{Soderblom}, D.~R. \& {Mayor}, M. 1993, \aj, 105, 226

\bibitem[{{Solomonoff}(1964{\natexlab{a}})}]{solomonoff64a}
{Solomonoff}, R. 1964{\natexlab{a}}, {Information and Control}, 7, 1

\bibitem[{{Solomonoff}(1964{\natexlab{b}})}]{solomonoff64b}
{Solomonoff}, R. 1964{\natexlab{b}}, {Information and Control}, 7, 224

\bibitem[{{Soubiran} {et~al.}(1990){Soubiran}, {Gomez}, {Arenou}, \&
  {Bougeard}}]{1990ebua.conf..407S}
{Soubiran}, C., {Gomez}, A.~E., {Arenou}, F., \& {Bougeard}, M.~L. 1990, in
  Errors, Bias and Uncertainties in Astronomy, ed. C.~{Jaschek} \&
  F.~{Murtagh}, 407--+

\bibitem[{{Starck} \& {Murtagh}(2006)}]{Starck06}
{Starck}, J.-L. \& {Murtagh}, F. 2006, {Astronomical Image and Data Analysis}
  ({Springer})

\bibitem[{{Stone}(1974)}]{stone74a}
{Stone}, M. 1974, Journal of the Royal Statistical Society. Series B
  (Methodological), 36, 111

\bibitem[{{Tollerud} {et~al.}(2008){Tollerud}, {Bullock}, {Strigari}, \&
  {Willman}}]{2008ApJ...688..277T}
{Tollerud}, E.~J., {Bullock}, J.~S., {Strigari}, L.~E., \& {Willman}, B. 2008,
  \apj, 688, 277

\bibitem[{{Tremaine}(1999)}]{1999MNRAS.307..877T}
{Tremaine}, S. 1999, \mnras, 307, 877

\bibitem[{{Tuominen} \& {Vilhu}(1979)}]{1979LIACo..22..355T}
{Tuominen}, I.~V. \& {Vilhu}, O. 1979, in Liege International Astrophysical
  Colloquia, Vol.~22, Liege International Astrophysical Colloquia, 355--360

\bibitem[{{Ueda} {et~al.}(1998){Ueda}, {Nakano}, {Ghahramani}, \&
  {Hinton}}]{Naonori1998}
{Ueda}, N., {Nakano}, R., {Ghahramani}, Z., \& {Hinton}, G.~E. 1998, in {Neural
  Networks for Signal Processing VIII, 1998. Proceedings of the 1998 IEEE
  Signal Processing Society Workshop}, 274

\bibitem[{{van Leeuwen}(2007{\natexlab{a}})}]{2007ASSL..250.....V}
{van Leeuwen}, F. 2007{\natexlab{a}}, Astrophysics and Space Science Library,
  Vol. 250, {Hipparcos, the New Reduction of the Raw Data} ({Springer})

\bibitem[{{van Leeuwen}(2007{\natexlab{b}})}]{2007A&A...474..653V}
{van Leeuwen}, F. 2007{\natexlab{b}}, \aap, 474, 653

\bibitem[{{Wallace}(2004)}]{wallacebook}
{Wallace}, C.~S. 2004, {Statistical and inductive inference by minimum message
  length} ({Springer})

\bibitem[{{Wallace} \& {Boulton}(1968)}]{wallace68a}
{Wallace}, C.~S. \& {Boulton}, D.~M. 1968, {Computer Journal}, 11, 185

\bibitem[{{Wallace} \& {Dowe}(1999)}]{Wallace99a}
{Wallace}, C.~S. \& {Dowe}, D.~L. 1999, {The Computer Journal}, 42, 270

\bibitem[{{Wallace} \& {Freeman}(1987)}]{wallace87a}
{Wallace}, C.~S. \& {Freeman}, P.~R. 1987, Journal of the Royal Statistical
  Society. Series B (Methodological), 49, 240

\bibitem[{{Williams} {et~al.}(2009){Williams}, {Freeman}, {Helmi}, \& {the RAVE
  collaboration}}]{2009IAUS..254..139W}
{Williams}, M.~E.~K., {Freeman}, K.~C., {Helmi}, A., \& {the RAVE
  collaboration}. 2009, in IAU Symposium, ed. J.~{Andersen},
  J.~{Bland-Hawthorn}, \& B.~{Nordstr{\"o}m}, Vol. 254, 139--144

\bibitem[{{Williams}(1971)}]{1971MNRAS.153..171W}
{Williams}, P.~M. 1971, \mnras, 153, 171

\bibitem[{{Wilson}(1966)}]{1966Sci...151.1487W}
{Wilson}, O.~C. 1966, Science, 151, 1487

\bibitem[{{Wilson}(1932)}]{1938AJ.....47...49R}
{Wilson}, R.~E. 1932, \aj, 42, 49

\bibitem[{{Windham} \& {Cutler}(1992)}]{Windham}
{Windham}, M.~P. \& {Cutler}, A. 1992, {J.~Am.~Stat.~Assoc.}, 87, 1188

\bibitem[{{Zador}(1963)}]{zador63}
{Zador}, P.~L. 1963, PhD thesis, Stanford U.

\bibitem[{{Zador}(1982)}]{zador82}
{Zador}, P.~L. 1982, {IEEE~Trans.~Information theory}, 28, 139

\bibitem[{{Zhao} {et~al.}(2009){Zhao}, {Zhao}, \& {Chen}}]{2009ApJ...692L.113Z}
{Zhao}, J., {Zhao}, G., \& {Chen}, Y. 2009, \apjl, 692, L113

\end{thebibliography}
\end{document}